\documentclass[preprint,12pt]{elsarticle}

\usepackage{amssymb}
\usepackage{amsmath}
\usepackage{hyperref}

\usepackage{booktabs}
\usepackage{xr} 
\usepackage[raggedrightboxes]{ragged2e}
\usepackage{pdflscape}
\usepackage{makecell}

\journal{International Journal of Digital Earth}

\begin{document}

\begin{frontmatter}



\title{AnywhereXR: On-the-fly 3D Environments as a Basis for Open Source Immersive Digital Twin Applications}

\author[label1,label2]{Alexander Klippel\corref{cor1}}
\affiliation[label1]{organization={Cultural Geography Research Group (GEO), Wageningen University \& Research},
             city={Wageningen},
             country={The Netherlands}}
\affiliation[label2]{organization={WANDER XR Experience Lab, Wageningen University \& Research},
             city={Wageningen},
             country={The Netherlands}}

\author[label2]{Bart Knuiman} %% Author name

%\affiliation{organization={},%Department and Organization
%          city={},
%           state={},
%           country={}}

\author[label4]{Jiayan Zhao} %% Author name

%\affiliation{organization={},%Department and Organization
%          city={},
%           state={},
%           country={}}

\author[label3]{Jan Oliver Wallgrün} %% Author name

\affiliation[label3]{organization={Independent Researcher},%Department and Organization
%          city={},
%           state={},
           country={Germany}
}

\author[label4]{Jascha Grübel} %% Author name

\affiliation[label4]{organization={Laboratory of Geo-information Science and Remote Sensing, Wageningen University \& Research},%Department and Organization
          city={Wageningen},
%           state={},
           country={The Netherlands}}

\cortext[cor1]{Corresponding author: alexander.klippel@wur.nl}

%% Abstract
\begin{abstract}
Visualization has long been fundamental to human communication and decision-making.
Today, we stand at the threshold of integrating veridical, high-fidelity visualizations into immersive digital environments, alongside digital twinning techniques.
This convergence heralds powerful tools for communication, co-design, and participatory decision-making.
Our paper delves into the development of lightweight open-source immersive digital twin visualisations, capitalizing on the evolution of immersive technologies, the wealth of spatial data available, and advancements in digital twinning.
Coined AnywhereXR, this approach ultimately seeks to democratize access to spatial information at a global scale.
Utilizing the Netherlands as our starting point, we envision expanding this methodology worldwide, leveraging open data and software to address pressing societal challenges across diverse domains.
\end{abstract}


\begin{graphicalabstract}
\centering
\includegraphics[width=1\linewidth]{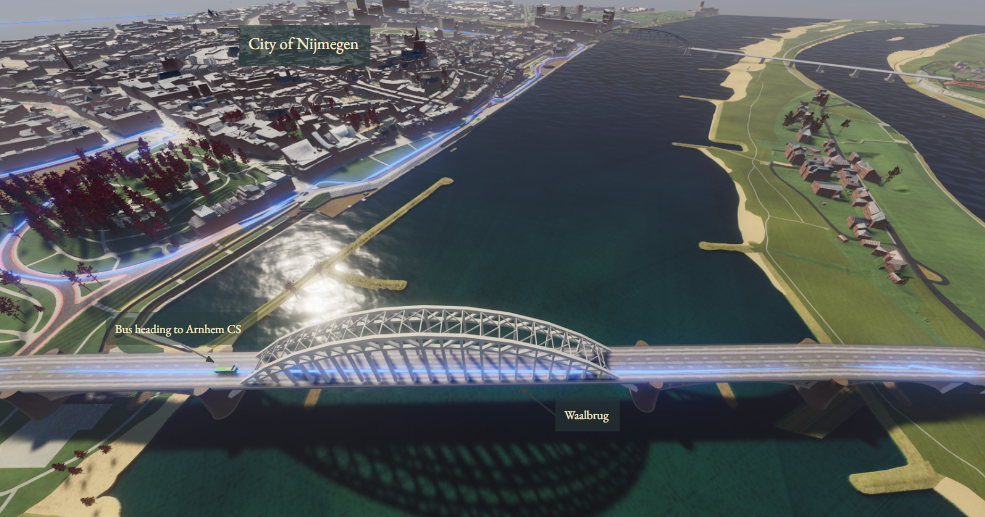}
\end{graphicalabstract}


\begin{highlights}
\item Lightweight creation of high-fidelity, object-based 3D environments on-the-fly based on public data sets
\item 3D environments suitable for creating interactive and adaptble immersive experiences through AR and VR
\item Connections to digital twin applications
\item Evaluation and case study
\item Enable futuring and co-design using immersive digital twins
\end{highlights}

%% Keywords
\begin{keyword}
digital twin
\sep immersive technologies
\sep extended realities
\sep public data
\sep visceralization
\sep digital humanism


%% or \MSC[2008] code \sep c<ode (2000 is the default)
\end{keyword}

\end{frontmatter}



\section{Introduction}

We inhabit a world inundated with data~\cite{sagiroglu.2013}, where the sheer volume of geospatial data collected daily is beyond imagination~\cite{Lee.2015,Liu.2022}.
This collection of data and the creation of a digital infrastructure across the world is seen as a key to addressing diverse planetary challenges~\cite{Hazeleger.2024,Loorbach.2017}.
Among these challenges are fostering innovation and establishing resilience among individuals, communities, and governments alike~\cite{BALOGUN2020101888}.
At the same time, we are still lacking efficient, human-centric methods~\cite{werthner2020vienna} to foster the development and design of more intuitive digital environments~\cite{Cosio.2023}.
More intuitive digital environments, for example, made possible through developments in immersive technologies, also referred to as eXtended Realities (XR), are envisioned to paradigmatically change how information is communicated and experienced, allowing humans to connect to each other and their (future/remote) environments at a more visceral level~\cite{7314296,rs14133095,Jamei.2017,Klippel.2021}. 

Human-centric interaction requires facilitating the ability to process information~\cite{SCAIFE1996185,Norman93things,Cheng.2013} as well as the experience of the information~\cite{Dourish2001where} towards a common goal that aligns with human values~\cite{prem2024principles} such as digital humanism~\cite{werthner2020vienna}.
In this context, the concept of \textit{external cognition} has been established to better understand the benefits of visual artifacts~\cite{SCAIFE1996185}.
By visually externalizing cognitive tasks, such as identifying the relationship between data points, cognitive capacity (also called cognitive load) is freed for more complex tasks. 
In recent publications, this concept has been expanded to not only visualize data, but also \emph{visceralize data}~\cite{9229242} using immersive technologies.
Data visceralization focuses on creating digital environments that allow for experiencing the data to gain even deeper insights. 
Such embodied experiences are fundamentally physical and social~\cite{Kirsh.2013} and can even produce measurable physiological reactions to changes in representations such as the future state of an environment~\cite{spielhofer2021physiological}.
Embodied interactions can shift the meaning-making from the designer to the users~\cite{Dourish2001where}.
Therefore, human-centric digital environments based on immersive technologies could facilitate the comprehension of data, enable smarter and more human decision making~\cite{werthner2020vienna}, and effectively communicate environmental or societal issues to both scientific and nonscientific communities (e.g.~\cite{Besancon.2021}).
Despite these opportunities and knowledge about the benefits of embodiment and data visceralization, however, we face a deficiency in designing engaging methods for information gathering, sharing, and decision-making, ranging from collaborative digital environments to digital serious games.

The aforementioned combination of externalization and embodiment~\cite{rs14133095} holds promise to address and potentially resolve some of the challenges associated with understanding and using large data sets and the complex systems they represent.
To best promote these opportunities, it becomes critical to establish a centralized and open infrastructure for data visualization and data visceralization to cultivate novel forms of digital environments for geospatial applications~\cite{schultes2022fair,aguilar2024experiments,gruebel2023design}.

The research questions we are addressing in this paper concern the extent to which publicly available geospatial data sources can serve as a foundation for data visceralization at scale, which technical challenges need to be resolved, and how are current developments evaluated by users.
To address these questions, we introduce AnywhereXR, a procedural 3D environment generator that allows for creating high-fidelity immersive environments for---eventually---any place on Earth; we exemplify the opportunities using the data-rich Netherlands as an example. The first version of AnywhereXR that we will discuss is conceptualized as a proof of concept to demonstrate what is possible relying on publicly accessible data and which workflows are necessary to efficiently use public data sources to create immersive 3D environments.
A central goal of our approach is to create immersive environments such that embodied experiences become possible at any scale including street-level and allowing for object-based interactions, adaptations, and simulations.

\begin{figure}[thb!]
\centering
\includegraphics[width=0.5\linewidth]{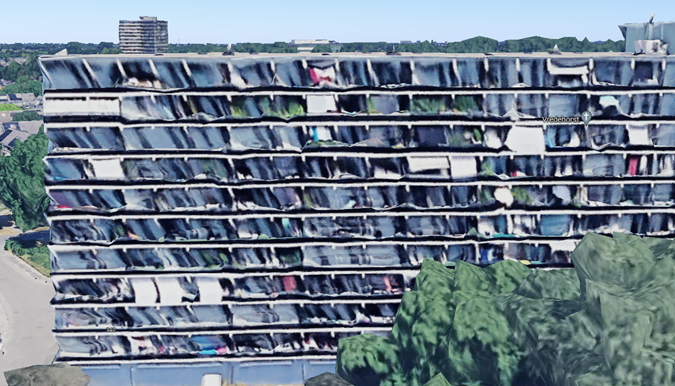}
\caption{Facades and trees from a 3D tile in Google Maps. 
(Source:~\cite{googlemaps_facade}).}
\label{fig:3dtile}
\end{figure}

While other approaches such as Google Maps, Google Street View, and Cesium exist, they are (semi)commercial products and largely based on voxel representations or meshes generated from point cloud data that do not provide sufficient fidelity at street-level (see Fig.~\ref{fig:3dtile}) or the option to interact with or modify the 3D environmental model at a level of entities meaningful to humans (such as trees, roads, or gateways.
In our advocated procedural approach, we are not aiming for competing with these products but for exploring an alternative based on open data that is representing the environment at the level of commonsense entities similar to what a 3D designer would use when manually modeling a given environment.
This object-based approach allows for applications that are otherwise not possible, such as adapting the model to illustrate potential future states or possible environmental design options, or adapting the model by modifying the state of objects based on real-time data feeds.

We provide a first expert-based evaluation of AnywhereXR that focuses on aspects of fidelity and completeness as well as a first case-study on integrating real-world data into an environmental model created with AnywhereXR.
While establishing a comprehensive comparison metric on technological (scalability, modularity, performance, etc.) as well as the perceptual (e.g., various aspects of fidelity) aspects for comparing different systems is a desirable goal for future research, we consider such an effort beyond the scope of the current paper, in particular because a fair comparison with the mentioned commercial products is a challenging task and the different areas of supported applications indicated above provide sufficient justification to investigate an object-based approach built on open data. 

We have structured the paper as follows: First, we provide necessarily short background information on immersive technologies, spatial data, digital twinning, and existing virtual representations of Earth (Section~\ref{sect:background}).
In Section \ref{sect:arch}, we detail the framework in which AnywhereXR operates and provide an example implementation.
In this section, we illustrate several of the technical solutions we implemented to incorporate publicly available spatial data into game engines as a basis for object-based immersive digital twins that can be experienced at any scale.
We give an overview of the data sources we used and discuss workflows, challenges, and solutions for several entities such as water bodies, buildings, or trees.
We then follow up with the two already mentioned evaluations.
The first one, detailed in Section \ref{sec:evaluation}, is an actual user evaluation using professional networks to elicit feedback and create an empirical assessment of the fidelity status of AnywhereXR.
The second evaluation is a case study (Section~\ref{sec:casestudy}) in which we consider AnywhereXR as the basis for an immersive digital twin application, demonstrating how live public transportation data can be fed into the AnywhereXR environment.
We conclude with a discussion for the wider XR and Digital Earth communities and an outlook on forthcoming applications.
We also briefly discuss opportunities for including generative Artificial Intelligence (AI) into the workflows to improve fidelity and discuss next steps (Sections~\ref{sec:gendisc}+\ref{sec:conclusions}).

\section{Background}
\label{sect:background}

We will first briefly review developments in immersive technologies for embodied experiences, the abundance of spatial data, the opportunity for digital twins to host such a setup, and existing virtual environments of Earth.
Last, we offer a synthesis, which we conceptualize as \emph{immersive digital twins}.

\subsection{Immersive Technologies}

Immersive technologies provide access to digital content via embodied experiences and replace sensory information from the physical world with digital content~\cite{Bowman.2007,Jerald.2016,Doerner.2022}.
As the human body is physically becoming part of the digital environment, the data representation is becoming accessible as a multi-sensory experience~\cite{Dourish2001where,Kirsh.2013,rs14133095}.
``Classic'' immersive technologies are virtual, augmented, and mixed realities, now jointly referred to as eXtended Realities (XR, but see~\cite{Flavian.2019,Rauschnabel.2022} for recent discussions about the terminology).

Immersive technologies are deeply intertwined with approaches in human-centric design~\cite{Slater.2016,Doerner.2022}.
Their inherent characteristics require prioritizing user experience and they are thought to provide intuitive interactions and accessibility.
Designed right, immersive experiences and technologies allow for leveraging human-centric designs and create digital environments that adapt to human needs rather than requiring user to conform to rigid systems.
By blurring the line between digital and physical (embodied) environments, designers have new opportunities to focus on ergonomics, cognitive load, and emotional impact of digital representations~\cite{Bailenson.2018}.
XR technologies have reached a comfortable level of maturity allowing for scaling accessibility into the realm of mass communication~\cite{9951054}.
Although universal adoption still remains elusive, the advancements resulting from the highly competitive landscape~\cite{9128969} are now yielding significant dividends for the deployment of XR.
We are currently witnessing the widespread adoption of immersive technologies among industry, scientists, and the general public.
In scientific circles, there has been a notable surge in articles focusing on XR and related topics~\cite{10.1145/3626517,Dane.inpress,Beck.2019,Kavanagh.2017}.
This trend suggests a growing acceptance of the technology beyond the realm of computer science, reflecting its development and increasing accessibility by several communities.
In many cases, the utilization of immersive technologies no longer requires specialized technical expertise.
Nonetheless, significant challenges remain in creating and designing content for immersive experiences.
Additionally, we need to further enhance the accessibility of immersive experiences and implement design choices validated through empirical research (\cite{Abeele.2021} and refer to Section \ref{sec:evaluation} for further details).

\subsection{Abundance of Spatial Data}

The increasing availability of (spatial) data is frequently hailed as a silver bullet to solve societal and planetary challenges~\cite{Kitchin2014,Hazeleger.2024,Bauer.2024}.
The idea is that with increasing data being available it becomes possible to make predictions, foresee trends, foster innovation, and create valuable insights.
At the same time, the abundance of data available has led to challenges~\cite{LIU2016134,Chen.2023,Liu.2022} and, from a human perspective, can be simply overwhelming~\cite{https://doi.org/10.1111/tops.12404}.
From the individual to governmental levels, we struggle with the flood of data and how to organize it~\cite{10.1145/1953122.1953140}.
This data deluge requires all actors to seek new ways for managing, analyzing, and creating meaningful insights from the available information~\cite{10.1145/3425709}.
Visualization has long been seen as a potent means to address the challenge of understanding data and especially large data sets~\cite{efe036efa6a945d68a809cdec9c1eb24,Keim.2001,Tang.2016,Dykes05:exploring}.
The Netherlands in particular offer a very rich set of openly available public spatial data ranging from elevation data to the build environment and the biosphere.
Therefore, we focus on the Netherlands to demonstrate AnywhereXR. However, in principle, this can be applied anywhere and we aim to expand the covered area.

\subsection{Digital Twins}

The term digital twin is not well defined and in many ways overused~\cite{Batty2024,rs14133095,Jones.2020}.
The early use of the term dates back to the early 2000s, describing the digital representation and modeling of a system and its functions~\cite{Jones.2020}.
Early definitions focus on the similitude~\cite{doi:10.2514/6.2012-1818} of represented system components and system processes.
These requirements were eventually relaxed due to their impracticality~\cite{Korenhof2021} and replaced with all kinds of digital representations and models that settle for eventual consistency~\cite{doi:10.1177/2399808318796416}.
For this paper, we highlight the following properties of digital twins:

\begin{itemize}
    \item A digital twin is a virtual representation of processes in the physical world (physical twin).
    These processes are not limited to physical processes but can also be social processes within the physical environment.
    \item A digital twin can represent natural, artificial processes in the physical world.
    \item A digital twin goes beyond a digital model by providing a connection or an option for a connection between the digital twin and the physical or social.
A connection may be incorporating data, potentially in real-time, from the physical twin into the digital twin, and vice versa.
    \item A digital twin should provide a process representation to allow for the exploration of current and past states and through simulation and prediction of future states of a physical system.
\end{itemize}

Another issue with digital twins is that they are often unique artifacts that cannot be generalized~\cite{rs14133095}.
By design, digital twins could be reproducible and reusable to facilitate the application in new areas, tasks, or cases.
To achieve this, it is required to move from designing individual digital twins to producing the infrastructure to generate digital twins on demand such as the Open Digital Twin Platform~\cite{swc2023odtp} and Digital Twins as a Service~\cite{10448890}.
To generate digital twins in context, a library of reusable components is required that facilitates tasks from data loading to data processing, data analysis, data management, and crucially data visualization.
As digital twins are naturally embedded in actual environments~\cite{pmc1016402022}, it becomes crucial to not only represent the physical twin digitally but also the surrounding environment.
Especially, if we consider the automated generation of practical digital twins, the infrastructure to generate credible 3D environments based on actual environments becomes a key criterion.

\subsection{Existing Virtual Environments of Earth}

Several services provide both online and offline virtual representations of Earth. There are two main approaches: using drone-based data or satellite-based data.
The first approach, used by platforms such as Google Maps, Cesium and ArcGIS Online, relies on drone-derived data to reconstruct 3D triangular meshes from LiDAR scans. This method, however, exhibits limitations in terms of street view quality and the ability to modify facades or soil types.

The second approach involves converting vector tiles (derived from remote sensing data) directly into triangular meshes, without taking the observer's viewpoint into account. While this method offers a streamlined way to generate virtual environments,  it proves inefficient at larger scales. The complexity of the triangular meshes remains uniformly high, irrespective of the observer's distance or perspective, leading to computational inefficiency. A publicly available implementation of this vector tile-based method for the Netherlands can be accessed via \href{https://www.pdok.nl/introductie/-/article/3d-basisvoorziening-1}{PDOK}.

We also want to emphasize again that none of these approaches is object-based in the sense that they would provide individual meshes or representations corresponding to the entities in which we conceptualize the world (for instance, individual meshes for each tree in the model). This makes these approaches highly problematic when it comes to modifying the models to simulate alternative states, to adapt the model based on real-world live information concerning the state of objects, and to realizing natural interactions.

\subsection{Synthesis}

These pivotal developments discussed above can be used as a basis to synthesize the concept of \emph{immersive digital twins} (IDTs).
We define IDTs as digital environments for human-data interaction~\cite{victorelli2020understanding} (but see~\cite{mab2021fused} for IDT's as physical environments for human-data interaction).
They combine digital twinning and immersive technologies to create experiential (embodied) access to data and simulations grounded in the physical (geospatial) world (see also~\cite{https://doi.org/10.1111/tgis.12932,Thuvander.2022,Keil.2021}).
IDTs underpin a paradigm shift for communicating and understanding data.
Whereas previously, data was communicated and illustrated in tables, figures, and maps to be read on paper or 2D screens, IDTs focus on ``entering'' the domain of the data and experiencing it physically with one's own body.
They also change both interpersonal communication and interaction in the context of data communication.
IDTs depart from conventional symbolic modes of communication and instead provide tangible and visceral, embodied experiences of past, current, or future states of a physical or social system.
IDTs potentially enable human-centric communication and facilitate meaningful experiences of information in large quantities of data.
The potential of IDTs and the different perspectives from which they are discussed have sparked the interest of corporations, governments, scientists, educators, and practitioners; although, naming conventions still have not converged~\cite{encyclopedia2010031}.

A complete and comprehensive discussion is beyond the scope of this paper but it is important to provide some key developments.
Early on, Norman \citep{Norman93things} noted the importance of tools for human cognition.
Knowledge in the head is not just connected to data in the physical world but also knowledge in the physical world.
Scaife and Rogers \citep{SCAIFE1996185} expanded on the concept and provided formalization as well as coining a term for it: external cognition.
They stress that cognitive processes seldom occur in isolation but should be seen as an integral part of a system with both internal and external representations.
While these authors are looking into a very broad range of artifacts and how they support and integrate with human cognition, the work by Clark \citep{Clark1989-CLAM} and others (e.g.,~\cite{Barsalou_1999}) stresses the importance of embodiment for cognition.
With the commercialization of XR, we are now in the position to design for embodied experiences at scale~\cite{Rodriguez.2021}.
However, whereas the hardware now allows for easy access to embodied experiences in XR, we are lacking the tools to easily create realistic, interactive environments of the physical world.
Through IDTs, we will be able to create reliable and scientific approaches to futuring, co-designing, and co-creation~\cite{rs14133095,Klippel.2021} and we demonstrate this with our approach AnywhereXR.

\section{Architecture and Design}
\label{sect:arch}

AnywhereXR is a lightweight procedural generator that creates object-based, high-fidelity immersive environments from open geospatial data based on a modular approach.
To keep AnywhereXR focused on visualisation, its key task is to query public data to visualise a specific geolocation.
AnywhereXR does not perform analysis but rather provides a backdrop to present spatial analysis or simulations in an environmental context with high fidelity.
Our proof-of-concept implementation for the Netherlands is realized as a software tool inside the Unity3D\footnote{\url{https://unity.com/}} game engine, commonly used for creating immersive digital twins or other XR/3D applications.
All steps for obtaining the required geospatial data and produce the 3D model can be conducted from the Unity editor application without the need for further tools or infrastructure.
In addition to the Unity-based implementation discussed in this paper, we are also working on R and Unreal implementations of the AnywhereXR approach (not discussed in this paper).

\begin{figure}[thb!]
\centering
\includegraphics[width=\columnwidth]{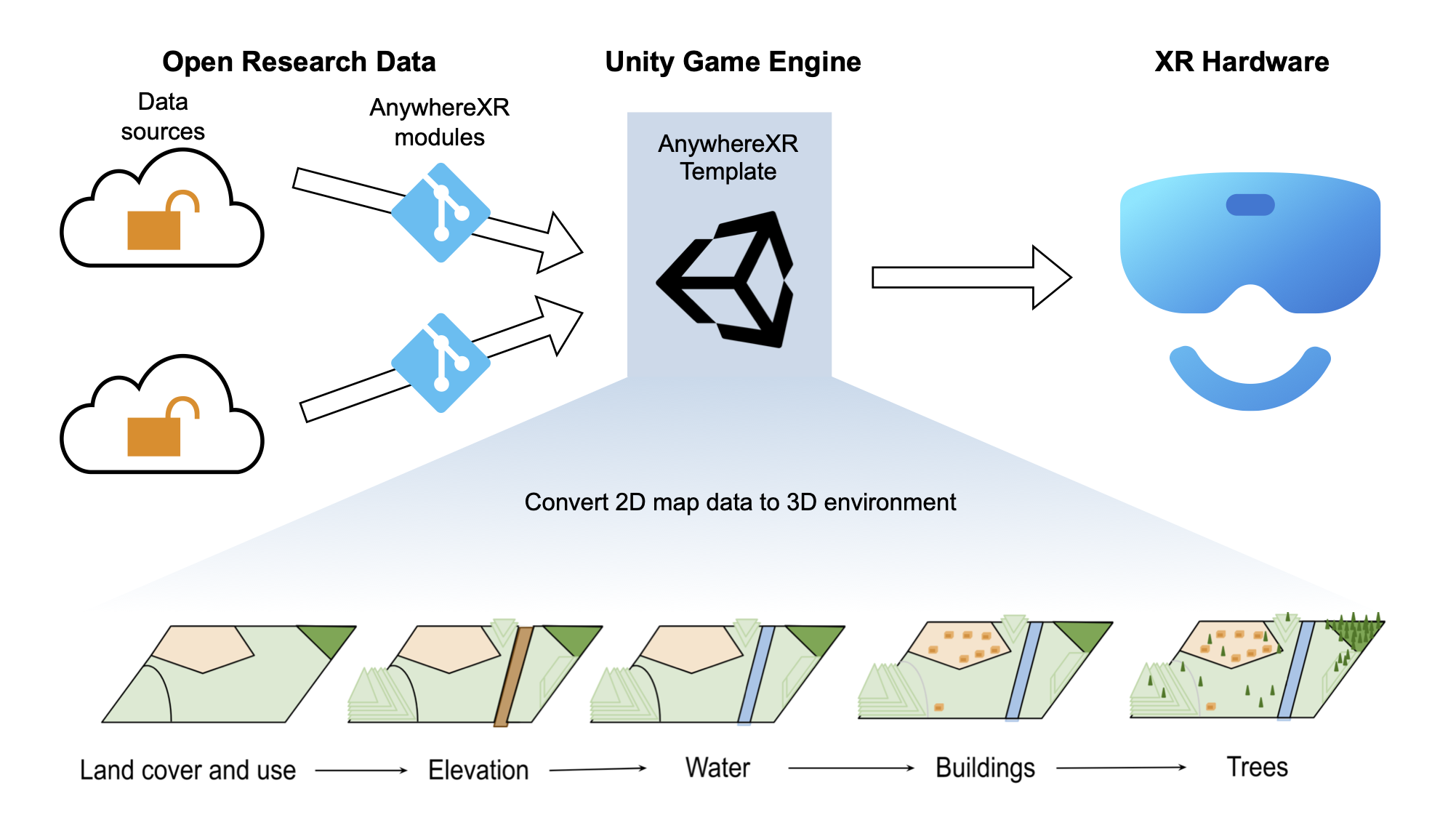}
\caption{Overview of workflow for AnywhereXR.
Data is loaded from open research data sources via AnywhereXR modules.
The AnywhereXR template selects modules and loads them from git repositories as unity packages.
2D maps are augmented throughout various steps to obtain high fidelity 3D environments, which can then be experienced in XR.
The augmentation steps at the bottom are shown from left to right: First, land cover and land use is rendered.
Second, elevation information is used to create terrain.
Third, water is rendered.
Fourth, buildings are placed.
Fifth, trees are rendered.}\label{fig:workflow1}
\end{figure}

AnywhereXR focuses on reproducing key characteristics of actual environments with the highest possible fidelity that public data offers.
It follows a modular design with separate unity packages for each stage dynamically loaded from git reposiories.
We present a basic AnywhereXR template focusing on five primary stages as shown in  Fig.~\ref{fig:workflow1}.
Each stage increments the depth and acuity of the virtual environment.
Logically, the stages progress as land cover/use generation, incorporating elevation, water generation, building generation, and tree generation.

To fully leverage the potential of IDTs in communication, education, decision-making, research, and stakeholder engagement, we advocate for the development of open-source, transparent, and replicable data and software environments~\cite{Barba.2022,schultes2022fair}, like AnywhereXR.
This aligns with ongoing initiatives focused on reproducible science (e.g.,~\cite{Baker.2016,aguilar2024experiments}), adherence to FAIR data standards~\cite{Wilkinson.2016}, and the empowerment of individuals through accessible data practices~\cite{Shadbolt.2012}.

\subsection{Data Sources}

To realize AnywhereXR, open source data for each stage must be available, which in many countries is still lacking.
Our proof of concept focuses on the Netherlands because it is not only densely populated but also surveyed and cataloged in great detail.
This also enables us to pursue an open research agenda by focusing on incorporating open data sources as much as possible.
For The Netherlands, we assembled the AnywhereXR pipeline from the following databases:
PDOK\footnote{\url{https://app.pdok.nl/vectortile-viewer/}}, 
AHN\footnote{\url{https://www.ahn.nl/}},
3DBAG\footnote{\url{https://3dbag.nl/en/viewer}}, and
Bomenregister\footnote{\url{https://bomenregister.nl}}.
To develop a better understanding of these databases and their potential for creating IDT applications, we explore their features and discuss their limitations below.

\subsubsection{PDOK}

PDOK stands for Publieke Dienstverlening Op de Kaart (EN: Public Service on the Map) and is the primary source of authoritative geoinformation and geodata sets from the Dutch government.
PDOK is a centralized database and platform for various data sets as well as services with the goal of facilitating public access, sharing, and use of geospatial data for a wide range of purposes.
PDOK offers detailed information on the cadaster, aerial imagery, administrative boundaries, hydrographic data, elevation models, and more.
The data is sourced from several authoritative agencies within the Dutch government such as the Kadaster and Land Registry.
Not all aspects of PDOK are included in our current approach, but thanks to the efforts of standardization (also in alignment with European efforts) and open-access, we are planning to further increase the use of PDOK.
For our proof-of-concept implementation, we focus mainly on the ground layers provided to define land cover and land use classes.

\subsubsection{AHN}

AHN stands for Actueel Hoogtebestand Nederland (EN: Current Height Model of the Netherlands).
AHN is a national database that provides very detailed information on elevation in the Netherlands.
It plays a vital role for the Netherlands given that a third of the country is below sea level and water is present almost everywhere.
A typical application of AHN is flood risk modeling.
The high accuracy is achieved through airborne LiDAR measurements which are regularly updated.
AHN also provides grid-based data such as 
Digital Terrain Models\footnote{\url{https://en.wikipedia.org/wiki/Digital_elevation_model}} (DTM) and Digital Surface Models (DSM).
In our proof-of-concept, we use AHN’s DTMs (see Section \ref{sec:elevation}).

\subsubsection{Bomenregister}

The Bomenregister (En: Tree Registry), is a Dutch database that provides centralized information about trees.
The Bomenregister is managed and maintained by municipalities, government agencies, and other organizations.
It is based on an initiative from 2014 by 
NEO\footnote{\url{https://www.neo.nl/}},
WUR\footnote{\url{https://www.wur.nl/en.htm}}, and 
GEODAN\footnote{\url{https://www.geodan.com/}}.
It contains detailed information of each individual tree such as their species, location, and at times historic information.
Unfortunately, the Bomenregister is not entirely free and only partially open access.
However, we include it here given the importance for fidelity and potential of this information being accessible.

\subsubsection{3D BAG}

3D BAG stands for 3D Basisregistratie Adressen en Gebouwen (EN: 3D Registry of Addresses and Buildings).
3D BAG is an extension of the BAG adding 3D information of all buildings in the Netherlands to the basic 2D information.
The database contains a number of building attributes such as geometric information of polygons that represent the outlines of individual buildings.
Additional semantic information provides insights into aspects of what buildings are used for and whether they are residential, commercial, industrial, or governmental.
The date of construction is also provided.
Data for the BAG are collected and maintained by municipalities, and then centralized and made available by the Dutch Kadaster.
To further extend the usage and usefulness of the BAG, the 3D version combines additional data sources such as AHN, BGT, and 
TOP10NL\footnote{\url{https://www.pdok.nl/introductie/-/article/basisregistratie-topografie-brt-topnl}}, into reliable 3D information.
The result is a nationwide data set with very precise information on the 3D structure of buildings in the Netherlands.
3D BAG\footnote{\url{https://3dbag.nl/en/viewer}} is maintained in collaboration with the TU Delft\footnote{\url{https://www.tudelft.nl/en/}} and is publicly accessible.

\subsection{AnywhereXR Processing Stages}

\begin{figure}[thbp!]
\centering
\includegraphics[width=\columnwidth]{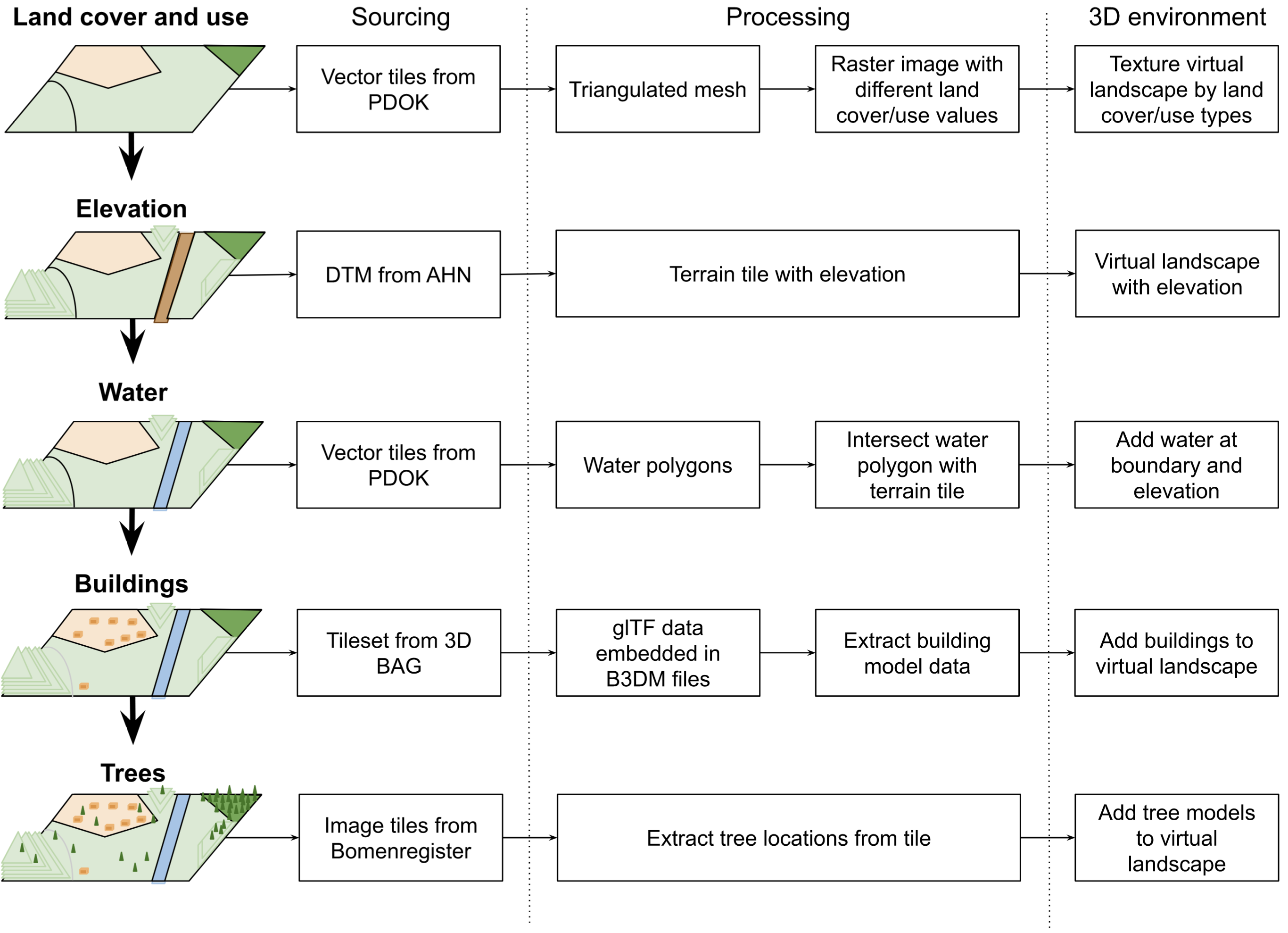}
\caption{Detailed Workflow for AnywhereXR.
AnywhereXR provides a foundation for immersive digital twins.
The workflow requires (public) data to instantiate immersive 3D environments.
The process is exemplified for the Netherlands, focusing on (largely) open data sources.
For each step, we describe the sourcing, processing and final 3D environment.
}\label{fig:workflow2}
\end{figure}

Each stage in AnywhereXR largely follows the same steps of sourcing the data, preprocessing it to extract relevant 3D information, and updating the 3D environment to increase fidelity.
The steps are shown in Fig.~\ref{fig:workflow2}.
We start with 2D map information, and we use a variety of steps to extrude 3D information.
Several sources of 3D information are noisy, incomplete, or projected into 2D, leaving us with missing information that needs to be reconstructed as the 3D environment is generated.
The main task of the AnywhereXR pipeline is to handle missing information and replace it with appropriate proxies.

\subsubsection{Land Cover and Use}

\begin{figure}[thbp!]
\centering
\includegraphics[width=\columnwidth]{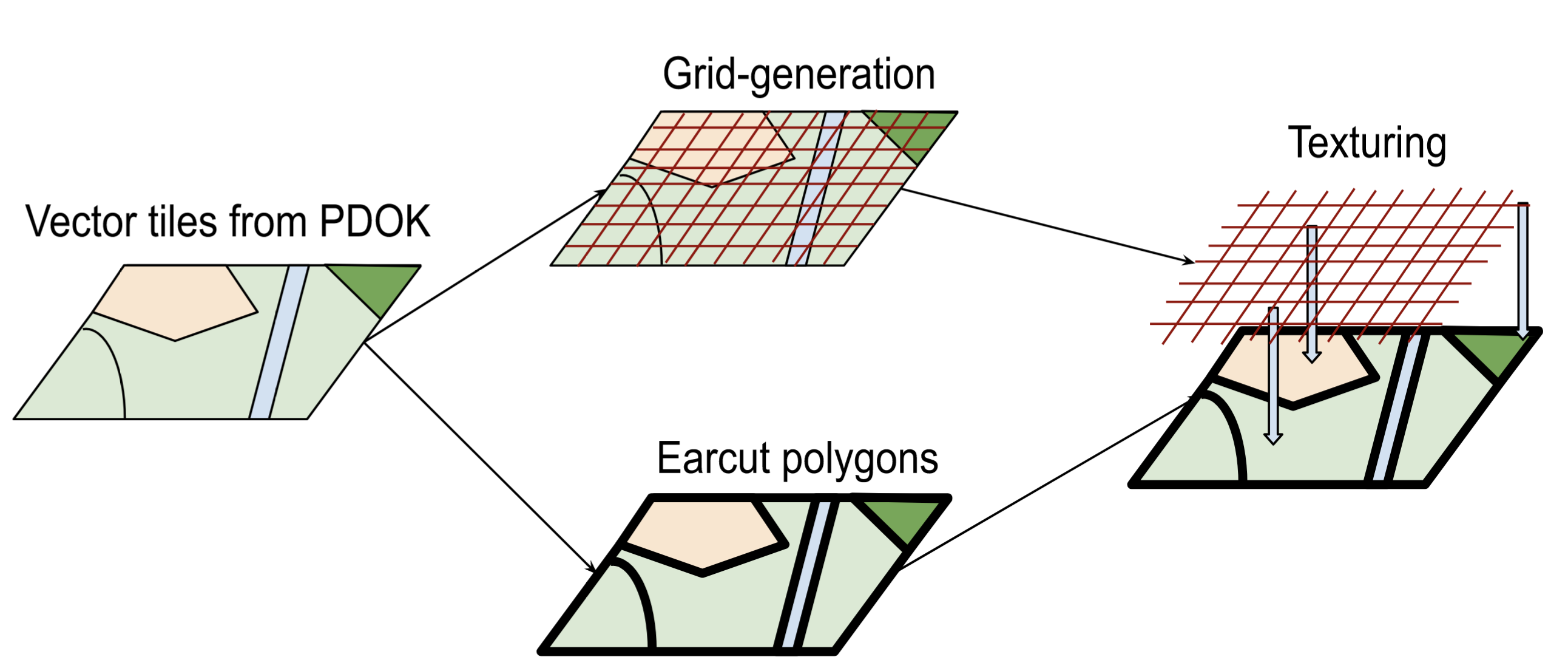}
\caption{The workflow for the generation of the environment surface and texture.
To reduce computational cost, a grid is sampled for texturing.
Earcut polygons are then used to produce high-fidelity land cover and land use textures on the grid.}\label{fig:workflow3}
\end{figure}

In the first stage, the surface of our virtual landscape is defined, labeled, and textured (see Fig.~\ref{fig:workflow3}). 
Our landscape generator uses vector tiles provided by PDOK that include land cover and land use information.
In PDOK vector tiles, natural and human-made features or regions (e.g., water, grass area, and streets) are represented by polygons that can be concave and contain holes.
Attributes of the polygons provide additional information.
As Unity does not support polygons with holes, we regularize the polygon by applying the 
Earcut\footnote{\url{https://github.com/mapbox/earcut}} algorithm to triangulate polygons with holes.
The resulting triangle meshes can be used inside Unity.

\begin{figure}[thbp!]
\centering
\includegraphics[height=5cm]{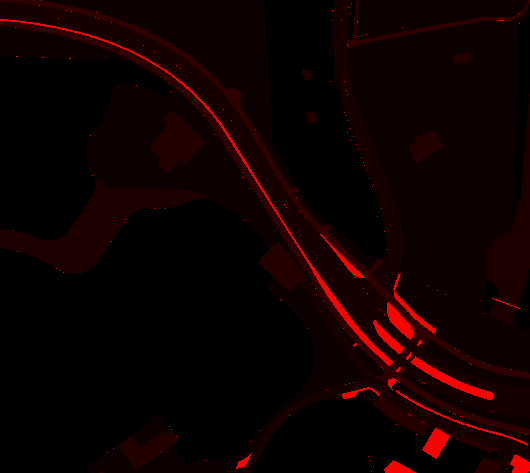}
\includegraphics[height=5cm]{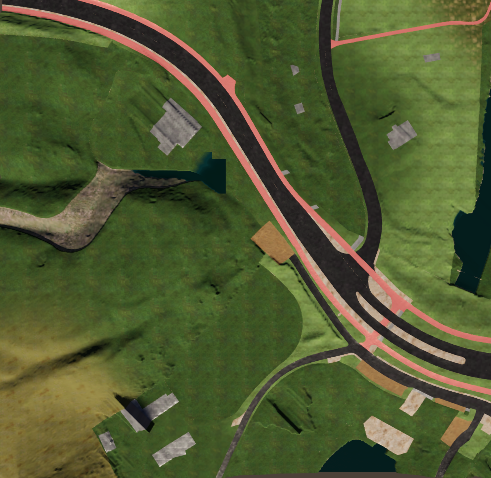}
\caption{Left: Raster image representing different land cover/use classes after sampling the vector tile from PDOK.
Right: Resulting terrain after applying texturing, elevation, and lighting.
}\label{fig:landscape}
\end{figure}

Computational efficiency also becomes an issue when texturing the polygons according to land cover and land use.
In principle, the information is available at the vertices of triangulated meshes.
However, this would produce inaccurate results in areas with a high density of height variation.
Whereas increasing the polygon count could overcome this, the resulting impacts on rendering performance would be high.
Instead, our solution employs a ray-tracing method to sample from the vector tile information using points in an N x N grid, recording interactions between the ray and any triangle.
The result is an N x N raster image attributed with spatial coordinates and a land cover/use value (e.g., 0 = water, 1 = grass, and 2 = cycling lane; Fig.~\ref{fig:landscape}, left).
This raster image is then used to assign a suitable texture to the polygons in the landscape grid (Fig.~\ref{fig:landscape}, right).

\subsubsection{Elevation}
\label{sec:elevation}

\begin{figure}[thbp!]
\centering
\includegraphics[width=\columnwidth]{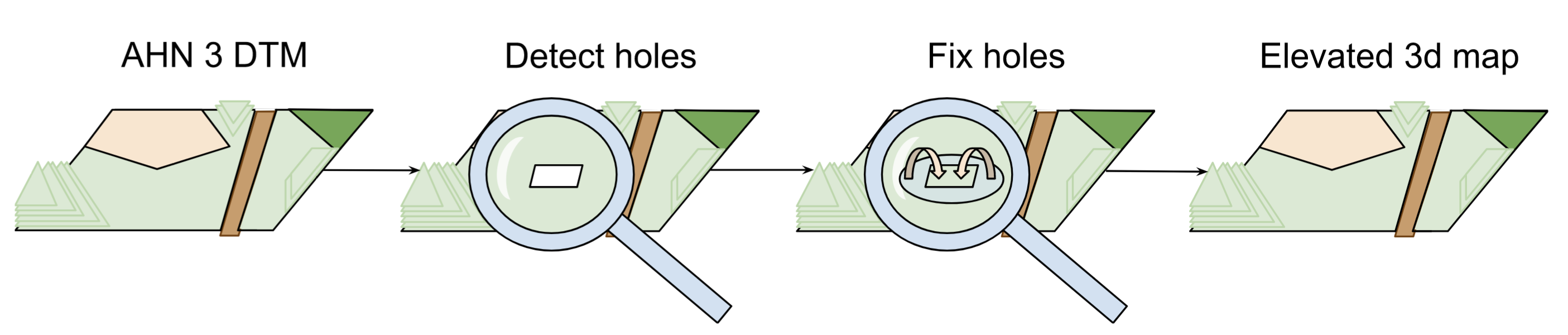}
\caption{AnywhereXR processes the DTM, fixes missing elevations, and adjusts the virtual environment.}\label{fig:workflow4}
\end{figure}

In the second stage, we augment the surface of the virtual environment with elevation information, see Fig.~\ref{fig:workflow4}.
A Digital Terrain Model (DTM) provides grid-based data concerning the varying elevations of the Earth's surface, specifically focusing on the natural terrain devoid of any superimposed structures or objects above street-level. 

\begin{figure}[thbp!]
\centering
\includegraphics[height=3.8cm]{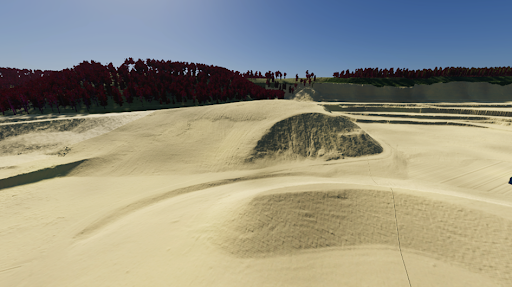}
\includegraphics[height=3.8cm]{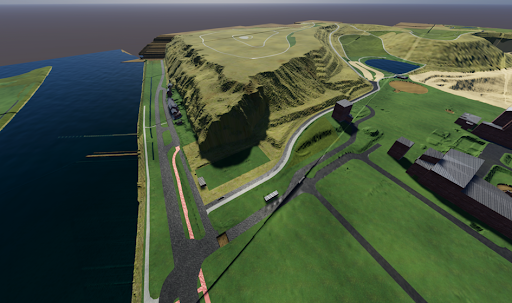}
\caption{Landscape created for Sint-Pietersberg south of Maastricht, illustrating the incorporation of high-resolution elevation data.
}\label{fig:terrain_examples}
\end{figure}

To add the elevation information to the virtual landscape, we are using the AHN 3 data with a precision of approximately 0.5 square meters (see Fig.~\ref{fig:terrain_examples} for examples of the generated terrain).
AHN 3 offers a higher resolution compared to its predecessors AHN 1 and AHN 2.
Whereas AHN 3 produces acceptable results of natural land surface environments, there are several limitations that need to be accounted for before applying it to our virtual environment.

\begin{figure}[thbp!]
\centering
\includegraphics[height=3.9cm]{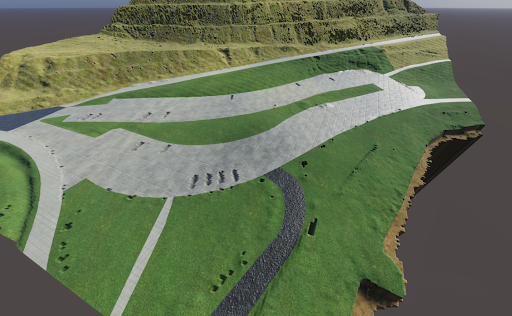}
\includegraphics[height=3.9cm]{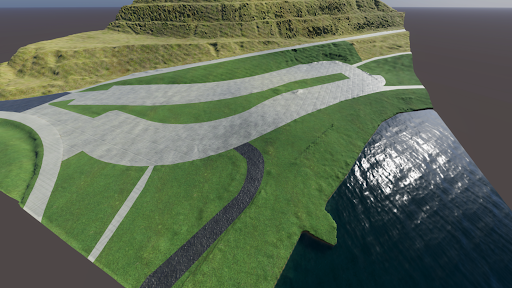}
\caption{Left: Raw, unfiltered terrain generated from AHN 3 data with frequent gaps and invalid water height.
Right: Result after post-filtering.
}\label{fig:terrain_postprocessing}
\end{figure}

The AHN 3 dataset is susceptible to producing erroneous readings in regions characterized by water bodies (see next section) or in areas with tall or complex buildings.
Typically, these regions produce missing data points which we mitigate with a post-filter mechanism.
The elevation at the missing data point is interpolated by leveraging valid samples in the neighborhood.
This post-filtering algorithm employs a circular search pattern centered around the missing data point.
From the neighborhood, several valid data points are extracted and weighted to fill the gap.
The process provides integrity and coherence of the generated terrain model (see Fig.~\ref{fig:terrain_postprocessing}).

\subsubsection{Water}

\begin{figure}[thbp!]
\centering
\includegraphics[width=\columnwidth]{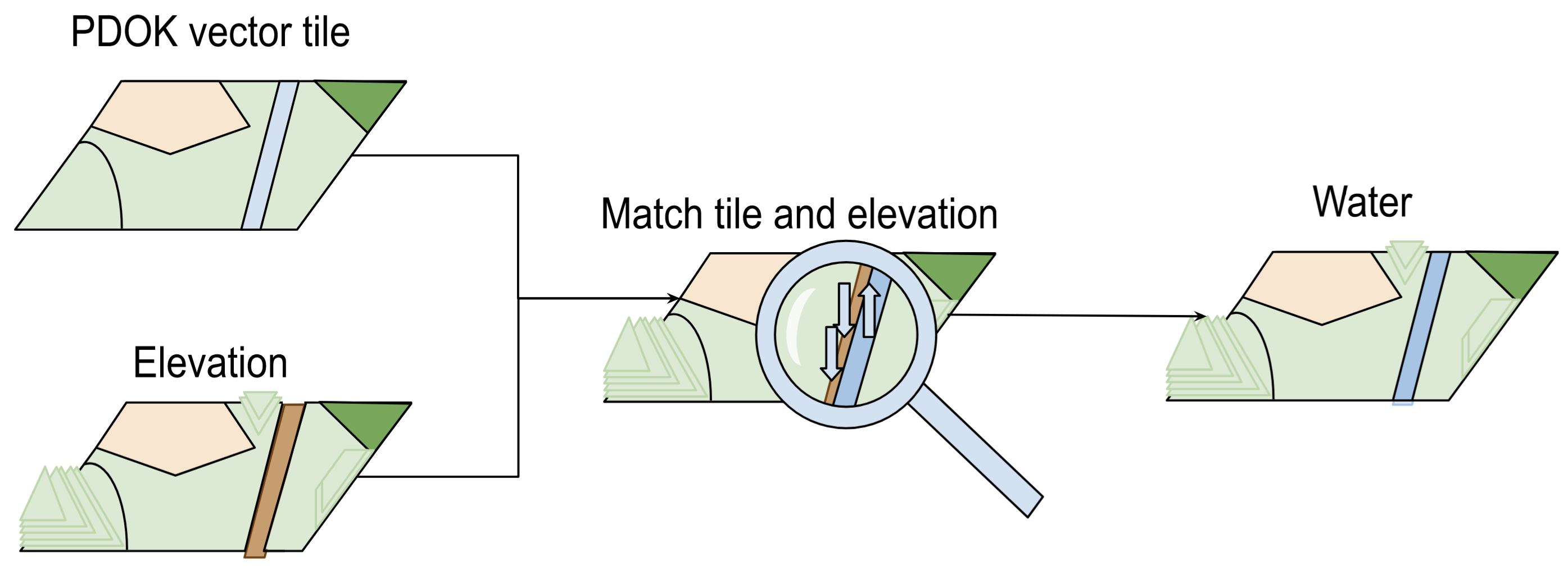}
\caption{Tile and elevation data are matched to obtain optimal boundaries.}\label{fig:workflow5}
\end{figure}

\begin{figure}[thbp!]
\centering
\includegraphics[height=3.5cm]{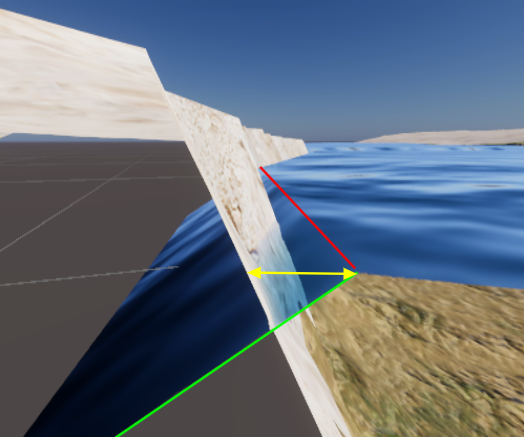}
\includegraphics[height=3.5cm]{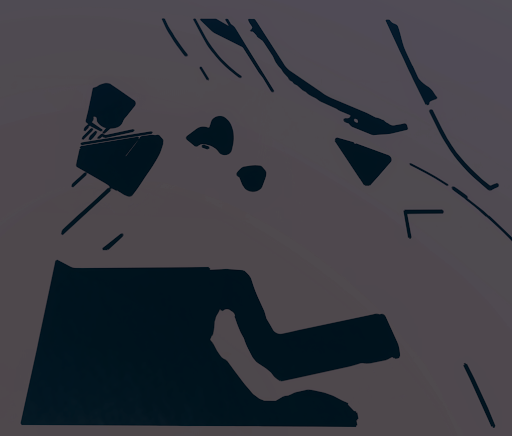}
\includegraphics[height=3.5cm]{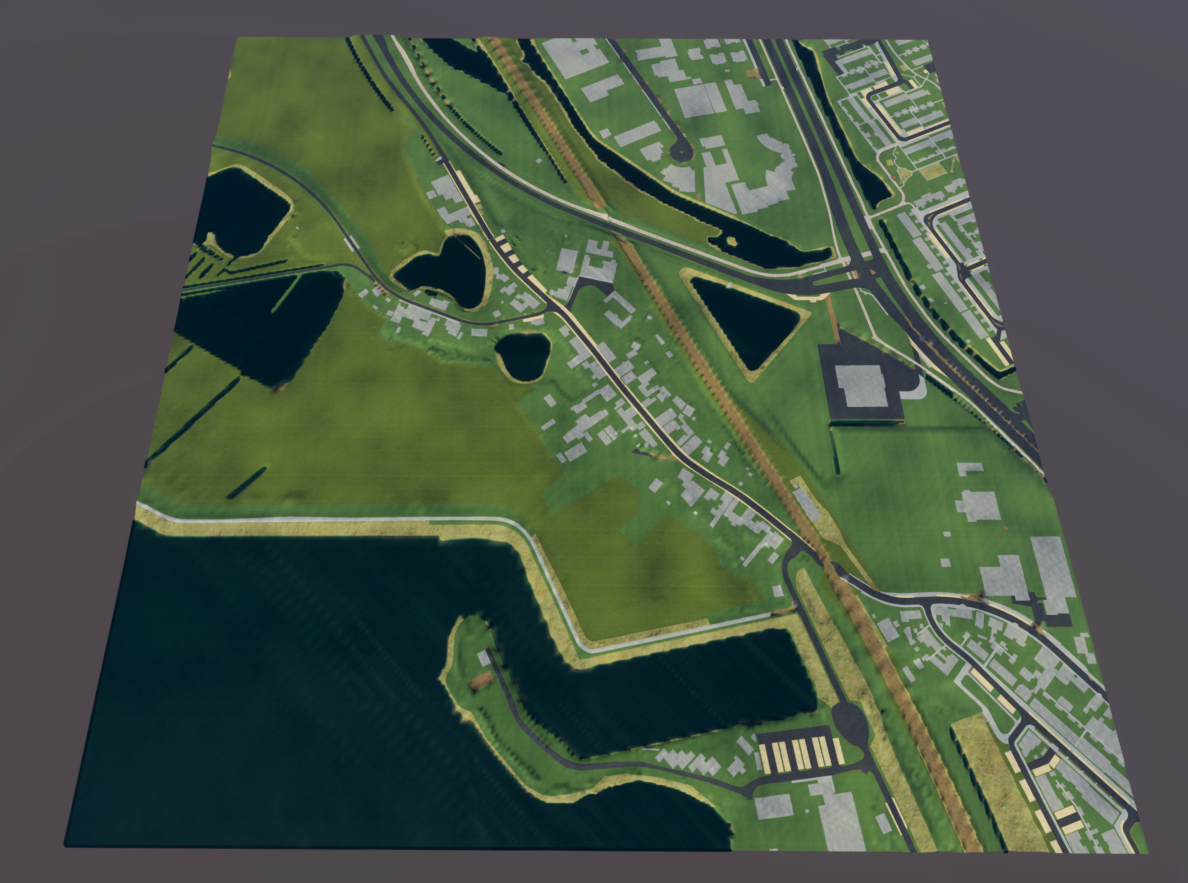}
\caption{Water generation.
Left: Gap (yellow) between terrain and water edge (red), and extruded water edge (green).
Middle: identified water bodies.
Right: Result of intersecting water with terrain.
}\label{fig:water_generation}
\end{figure}

Leveraging open geospatial data for 3D landscape generation presents challenges, particularly when incorporating water features.
While vector tiles from open repositories provide detailed information on lakes, rivers, and land cover (soil types, infrastructure, roads), their inherent 2D nature presents a hurdle for creating realistic 3D landscapes with elevation data.
The primary challenge lies in reconciling the elevation data with the vector tile information (Fig.~\ref{fig:workflow5}).
Elevation data, obtained from AHN in our case, will often not align perfectly with water boundaries in the vector tiles (Fig.~\ref{fig:water_generation}, left).
This mismatch arises because elevation data within water regions is often unreliable, making it unusable for determining water height.
Consequently, alternative methods are needed to establish water elevation.
Sampling the surrounding terrain offers a potential solution, but discrepancies between water boundaries and elevation data can lead to conflicts.

The advance of realism in landscape textures is contrasted with the difficulty of rendering translucent water.
The inclusion of water within the terrain grid presents a challenge due to the inherent limitations of the rendering method in case water would be simply displayed as a blue texture.
Unlike other terrain elements such as grass and sand, which seamlessly integrate into the grid with distinct textures, such a representation of water lacks the dynamic properties necessary for a realistic representation.
Rather than accurately reflecting its surroundings, refracting light, and responding to specular highlights, water would be reduced to a static, uniform blue patch on the ground.

More complex water shaders are available but in combination with elevation data, the visual effects can be erroneous.
Maps display all water within the plane, whereas in reality water sits on top of elevations.
Elevation data in AHN is often faulty at water boundaries as reflection and refraction produce discrepancies.
Therefore, the water polygons and land polygons would not match if simply displayed.
We attempted several solutions to this problem.

In an initial attempt to create realistic water surface meshes, we queried the control texture to determine the presence of water.
The returned value per pixel of this texture indicates a land cover type with “0” being water. 
Once identified, these water cells are aggregated and merged to ensure a smooth transition between water and terrain, thereby minimizing visual discrepancies along the edges (Fig.~\ref{fig:water_generation}, middle).
However, using this method, there exists a trade-off between the size of the water cells and rendering performance.
On the one hand, a very small cell size is desirable to result in smooth transitions and reduce the likelihood of inconsistencies due to overlap between neighboring water bodies.
On the other hand, this substantially increases the number of polygons that need to be rendered, particularly in areas where the water does not intersect with the surrounding terrain.
Consequently, this leads to a significant performance overhead.
In practice, this initial approach often necessitated manual adjustments to water elevation levels to achieve satisfactory outcomes.
However, this contradicts the objective of an automated generator.

\begin{figure}[thbp!]
\centering
\includegraphics[height=3.5cm]{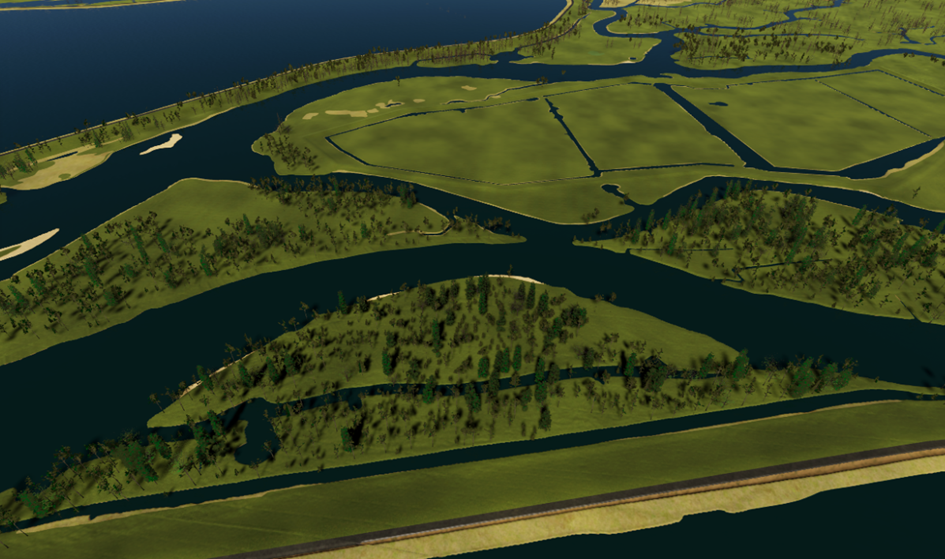}
\includegraphics[height=3.7cm]{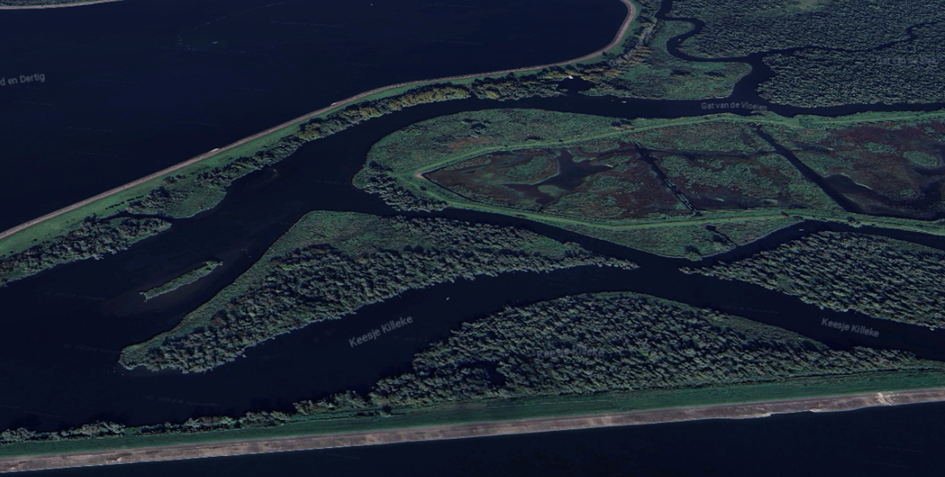}
\caption{Left: Biesbosch area automatically generated water representation.
Right: The same area in Google Maps as a comparison (Source:~\cite{googlemaps}).}
\label{fig:water_result}
\end{figure}

In response to this challenge, we adopted a different strategy by directly extracting polygons from vector tiles and then identifying the intersection between these water polygons and terrain tile boundaries (see Fig.~\ref{fig:water_generation}, right).
This approach facilitates more precise water elevation determination despite persistent discrepancies between AHN data and water boundaries that would otherwise not only impede accurate elevation calculations but also result in gaps along the water's edge, where it fails to seamlessly blend with the shoreline. 
Using the Biesbosch area in the Netherlands as an example, Fig.~\ref{fig:water_result} illustrates the satisfying water generation results we achieve with this approach.

\subsubsection{Buildings}

\begin{figure}[thbp!]
\centering
\includegraphics[width=\columnwidth]{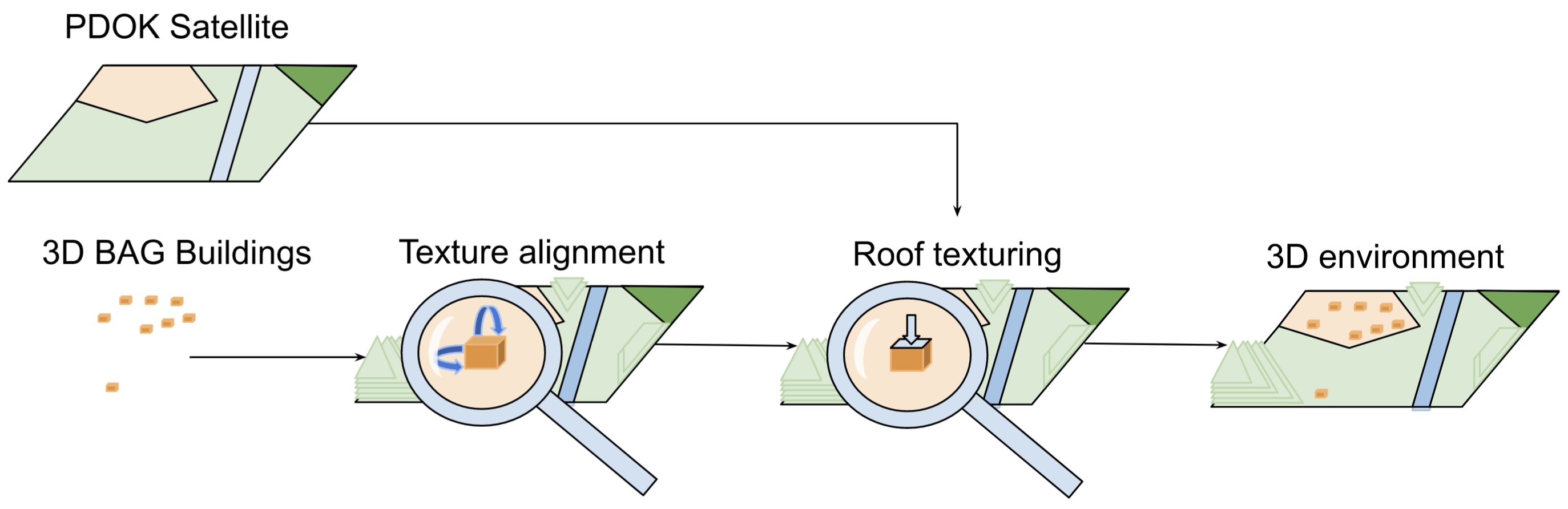}
\caption{AnywhereXR instantiates buildings from 3D BAG and preprocesses the textures and satellite images of roofs to produce a high-fidelity digital twin.
}\label{fig:workflow6}
\end{figure}

3D BAG offers an extensive repository of structural information encompassing 3D information about nearly every building across the Netherlands in the form of point clouds.
It offers the option to access the data for a specific region and download the result via the 3D BAG web viewer in the form of a \texttt{tileset.json} file.
This tileset structure uses a quadtree organization to efficiently manage references to .b3dm files with geometric building information.
The Batched 3D Models or .b3dm format is a standard for storing and streaming large-scale city models~\cite{10.1145/2945292.2945312}.
The files contain embedded glTF data that encapsulates batch tables with the properties of multiple objects as vertex attributes~\cite{Khronos.org}.
Our model generator uses these tables to obtain vertex information for individual buildings and translate this information into Unity meshes (Fig.~\ref{fig:workflow6}).

\begin{figure}[thbp!]
\centering
\includegraphics[height=3.75cm]{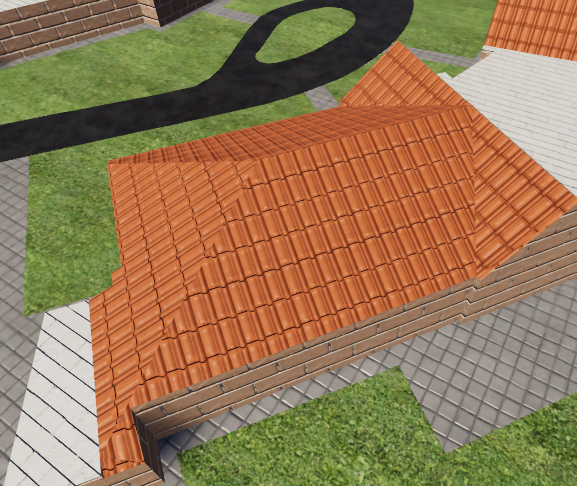}
\includegraphics[height=3.75cm]{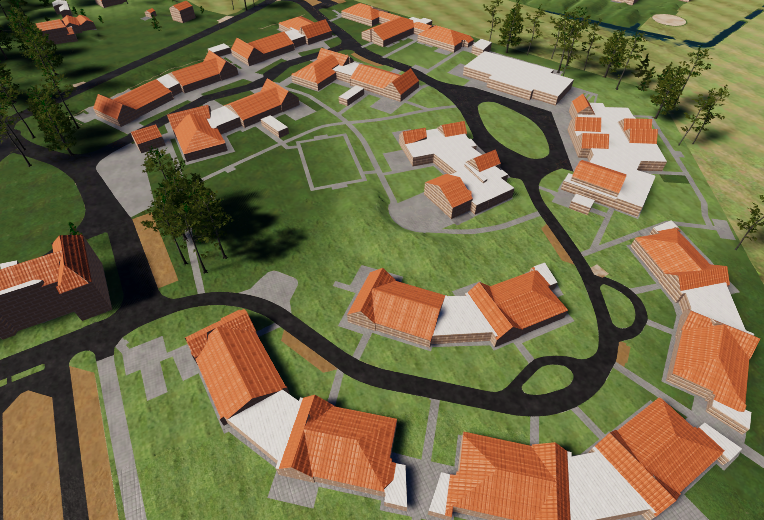}
\caption{Left: Texture mapping for building facades. Right: Group of houses for an area in Scheveningen, The Netherlands.
}\label{fig:buildings}
\end{figure}

\paragraph{Texture mapping} The .b3dm files do not include the texture coordinates for individual buildings.
For walls and angled roofs that are not parallel to the ground, the texture can be rotated and rescaled to align it with the specified plane (Fig.~\ref{fig:buildings}, left).
For flat roofs, applying the texture is even more challenging due to the lack of orientation information.
In order to determine where to place textures, a series of vector operations are performed that map positions (horizontal and vertical) and extents (width and height) of textures onto given building facades (i.e., UV mapping; see~\cite{Flavell2010} for explanation).
Fig.~\ref{fig:buildings}, right, shows an example result of our building generation approach in Unity.

\begin{figure}[thbp!]
\centering
\includegraphics[height=3.55cm]{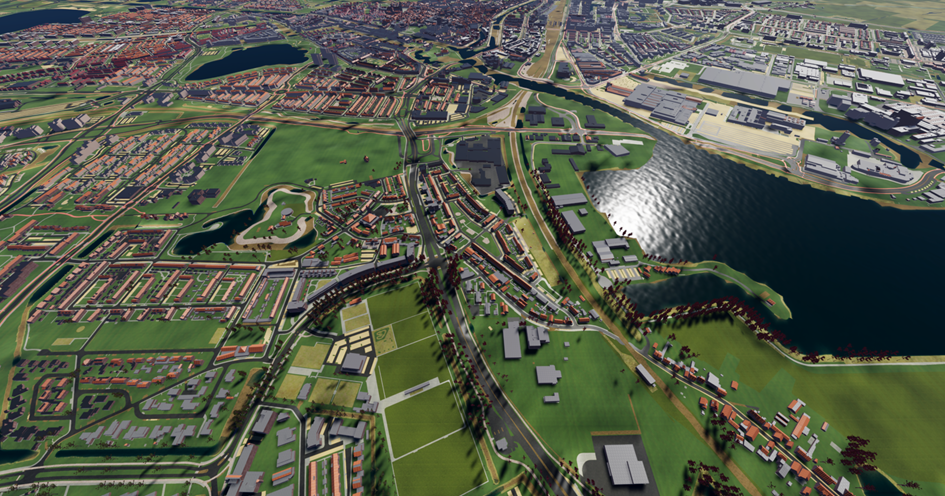}
\includegraphics[height=3.55cm]{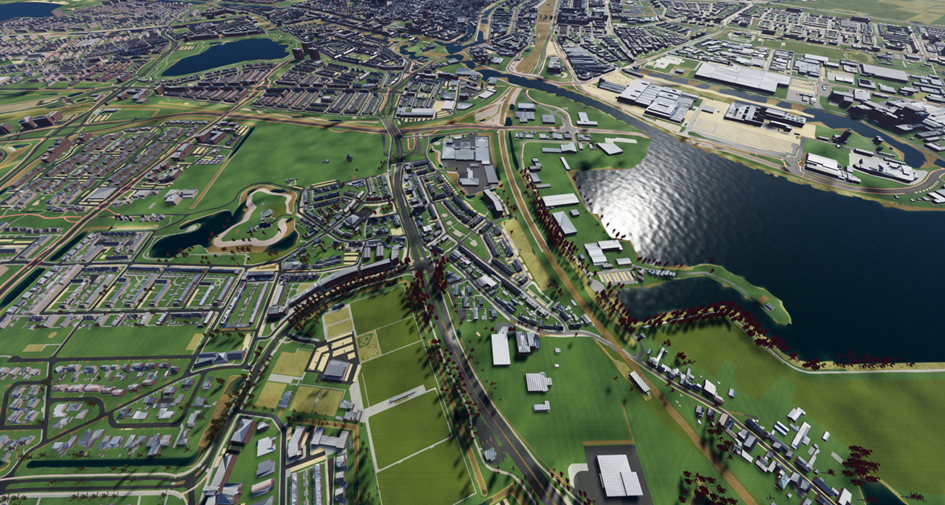}
\caption{Den Bosch with pseudo-random roof colors (left) and roof colors derived from aerial imagery (right).}\label{fig:roofcolor}
\end{figure}

\paragraph{Roof coloring} After initially using randomly selected colors for the roof polygons, we made a move towards greater realism by extracting the color for each polygon from aerial imagery provided through PDOK.
It is important to note, though, that this aerial imagery comes with several caveats.
These include color variations due to factors such as time of year, sun position, satellite position, presence of debris or solar panels on roofs, accumulation of dirt, and numerous other elements that can complicate the selection of an accurate color representation.
Fig.~\ref{fig:roofcolor} shows the same area using the initial approach (left) and improved version (right).

\subsubsection{Trees}

\begin{figure}[thbp!]
\centering
\includegraphics[width=\columnwidth]{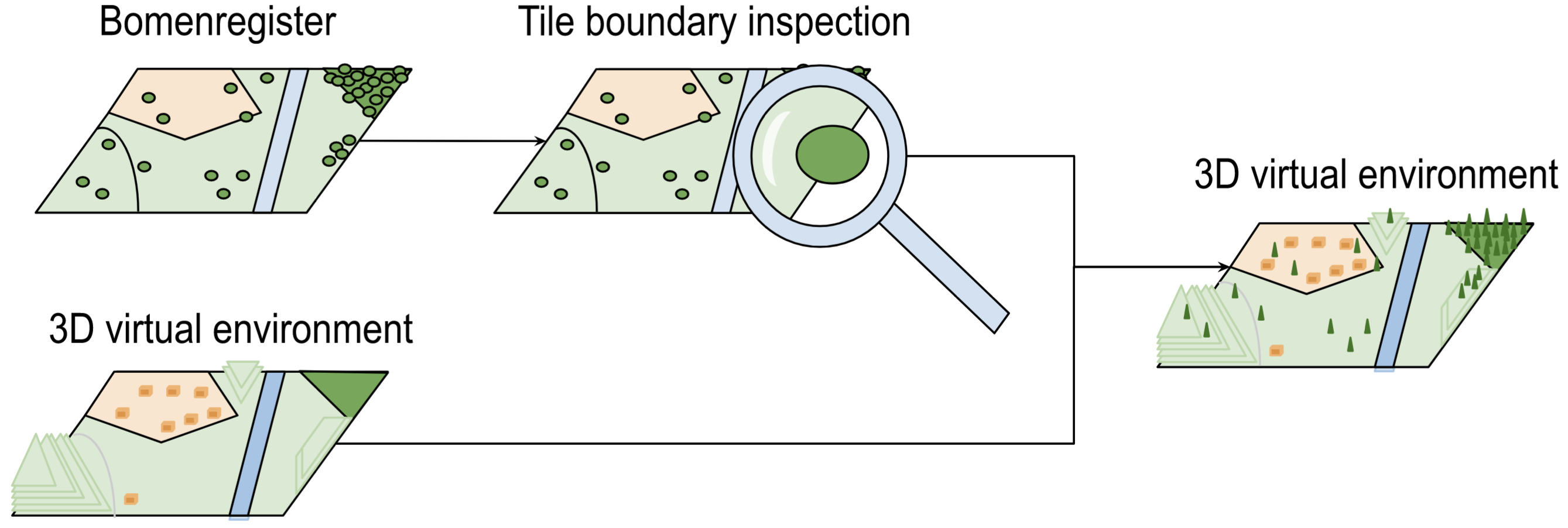}
\caption{AnywhereXR uses the Bomenregister to instantiate trees in the virtual environment. The boundaries of tiles are inspected to remove duplicate trees present in multiple tiles.
}\label{fig:workflow7}
\end{figure}

\begin{figure}[thbp!]
\centering
\includegraphics[height=3.9cm]{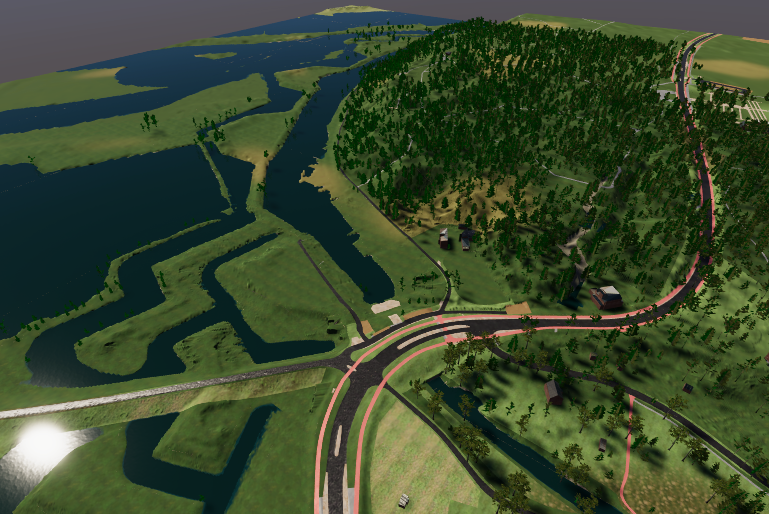}
\includegraphics[height=3.9cm]{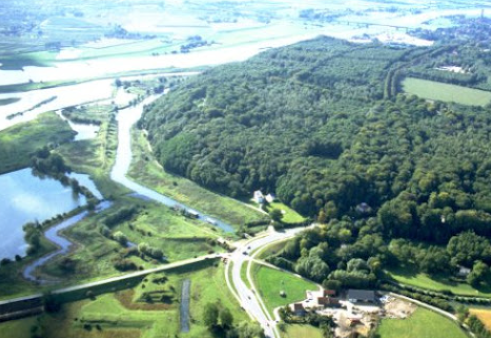}
\caption{Left: Generated landscape representation for Grebbeberg, Rhenen, The Netherlands, with tree models placed at the positions extracted from Bomenregister image tiles. Right: Actual photo of the area as a comparison.
}\label{fig:trees}
\end{figure}

After water bodies and buildings, trees are the next important discrete landscape feature in our model creation process (Fig.~\ref{fig:workflow7}).
Our landscape generator obtains information on tree distributions from the Bomenregister web portal.
A downloaded dataset contains image tiles that show tree crowns as distinct regions with color values that represent the tree size.
This data is processed by applying a flood-fill algorithm to identify individual trees and determine their position by taking the centroid of the corresponding region in the image.
In the second processing step, our landscape generator deals with trees located at the boundaries between adjacent tiles to create cross-references between areas in different tiles that belong to the same tree.
The final tree entities and their locations are then used to place 3D models of trees on the terrain.
Fig.~\ref{fig:trees} shows the produced result obtained for the Grebbeberg, Rhenen, and a photo of the region for comparison.

\section{Evaluation}
\label{sec:evaluation}

Evaluation was conducted internally following an iterative design approach and with expert users in an online survey.

\subsection{Iterative design approach}
We have been discussing an automated workflow for creating 3D environments for immersive digital twins from public and semi-public data.
Our approach first focused on solving many of the technical challenges.
The evaluation therefore followed an iterative design approach~\cite{10.1109/2.241424} where developments are assessed and improvements are discussed by the experts in our team.
The guiding design principle for the overall approach is inspired by the visual Turing test~\cite{shan2013visual};
without specifying a context of use, we aim for a replication of the physical environment in a veridical, high visual-fidelity way.
Following the ideas behind the original Turing test, the ultimate, long-term goal is to create and design digital environments indistinguishable from physical environments.
Currently, the availability of publicly available data for the veridical representation of an environment is still too sparse.
However, the developments in rendering technology and immersive technologies point towards the eventual possibility of creating environments that can pass the visual Turing test, assuming the data is available.
As of today, we make use of the available data to provide a level of fidelity suitable for many applications or the basis for further developments.

%One interesting aspect of our approach is that the visual Turing test would not only work for the present but also for future (or past) states of the environment.
%Our IDT framework relies on data from databases and automates the creation of 3D virtual environments that can be experienced through immersive technologies.
%We are not discussing here how such databases are filled with data and information as this is beyond the scope of the paper.
%By connecting to worldwide digital twin efforts~\cite{Boschert.2016,Jones.2020} and our own work on visceralizing data on forest futures [anonymized] and visualization of historical data [anonymized], we foresee tremendous potential for making such data more accessible.
% last two references for jiawei's papers

\subsection{Survey evaluation}
In addition to the iterative design approach, we performed an evaluation of different environments (e.g., forest, coast, urban) comparing images captured in Google Earth or taken of the physical world to images taken from our AnywhereXR environments. We did so using an anonymous survey widely shared with different relevant communities. The details and outcomes of the survey are as follows.

\subsubsection{Materials} 
\begin{figure}[thbp!]
\centering
\includegraphics[height=3.35cm]{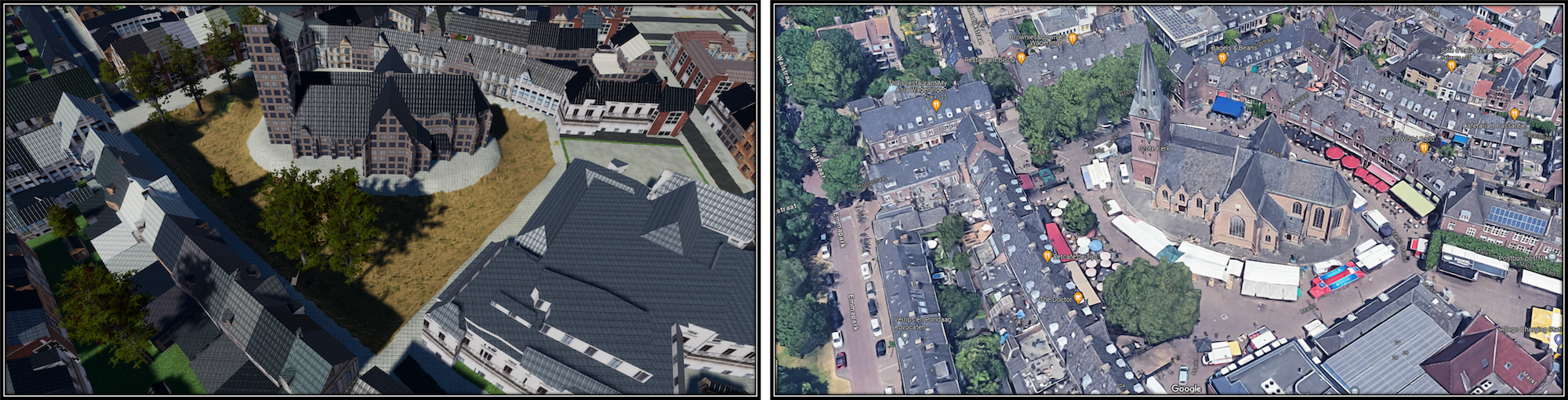}
\includegraphics[height=3.45cm]{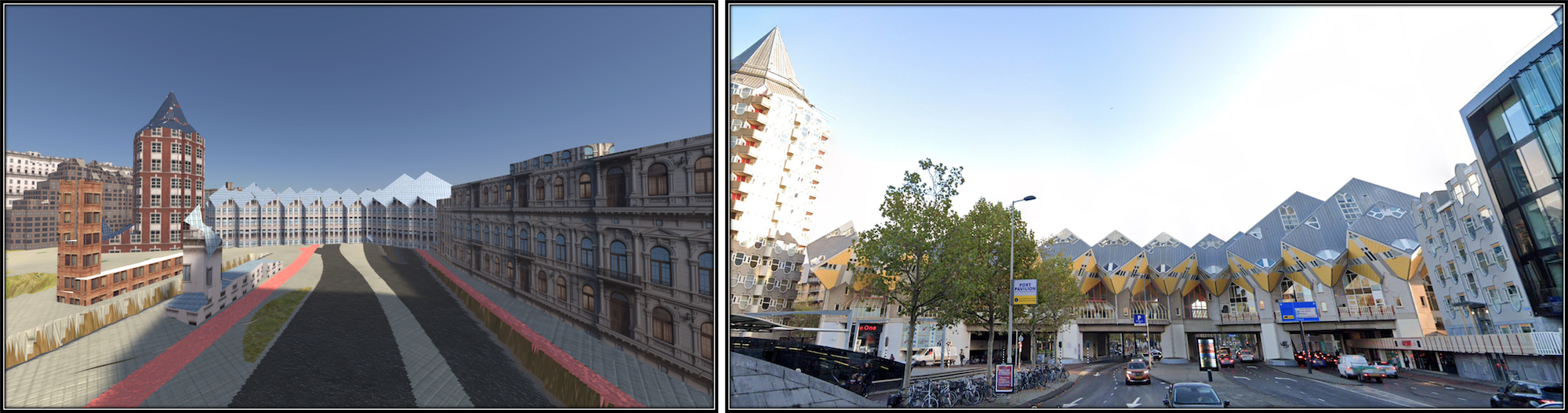}
\includegraphics[height=3.0cm]{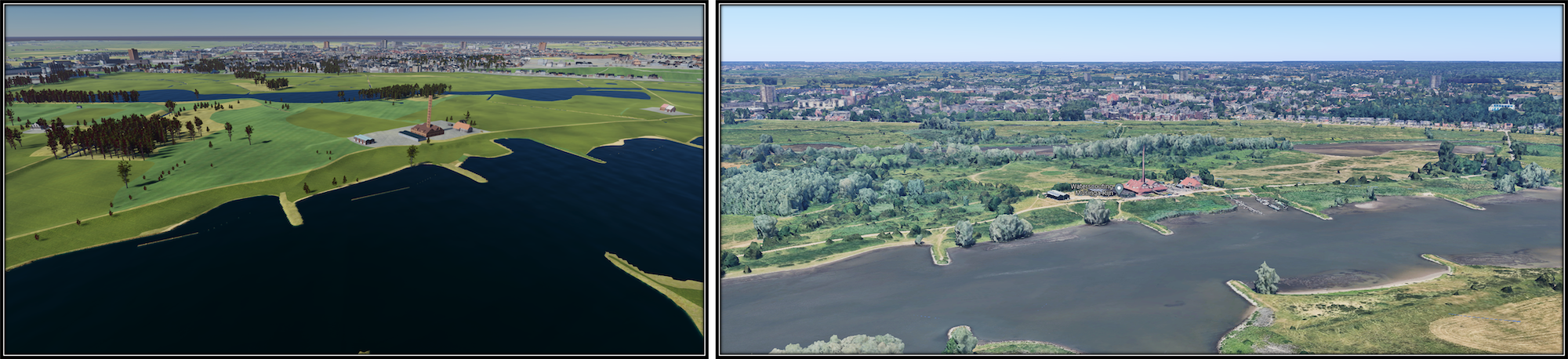}
\includegraphics[height=3.25cm]{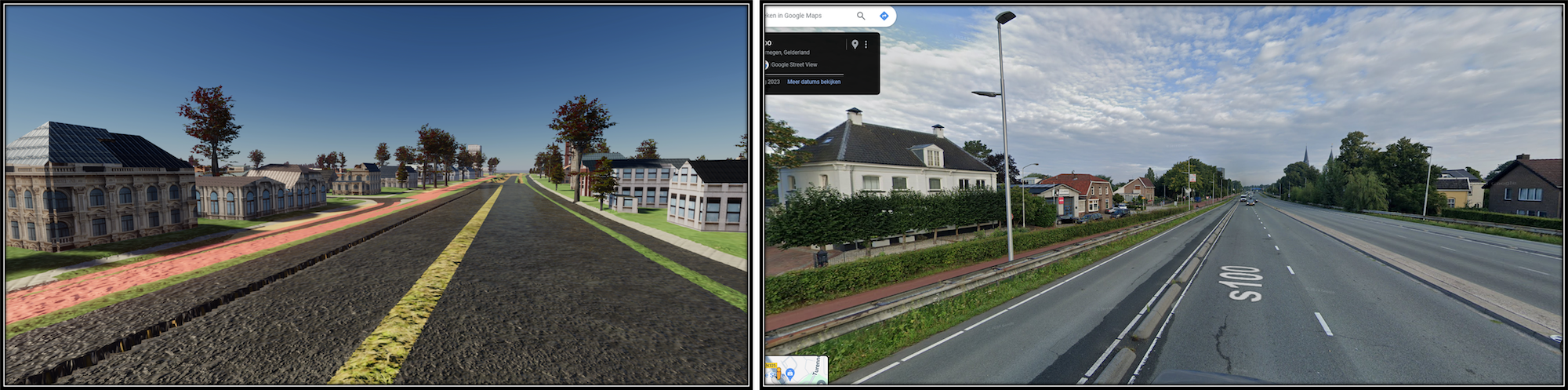}
\caption{Depicted are examples of the comparisons of AnywhereXR (left) and benchmark worlds (right, generated from Google Earth or real-world camera footage) as used in the survey.
While not always identical, the perspectives and viewing angles were matched as much as possible.
Top image pair received the highes ranking; second from the top the lowest.}\label{fig:examplesSurvey}
\end{figure}

To assess the level of perceived visual fidelity in our automated AnywhereXR application, we selected 10 locations throughout the Netherlands from urban as well as suburban, natural and coastal areas.
For these 10 locations, we created corresponding image pairs of the physical world and the virtual world.
Figure \ref{fig:examplesSurvey} provides examples of the image pairs.
Some of the images were taken from a birds-eye perspective, some at street-level.
We tried to match perspectives and visible area as much as possible.

\subsubsection{Participants} The survey is based on a convenience sample and was distributed widely via relevant email lists such as 3DUI, AGILE, ICACogVIS, ISPRS, and LinkedIn.
There was no restriction on who could participate and participation was entirely voluntary with no further incentives provided.
A total of 78 people responded to the invitation, from this, 29 participants filled out the questionnaire completely. 
Average age of participants was 43 years (SD=12.64). Nine participants were female (20 male). 
Almost all participants had an academic background with the vast majority being in geo-information or computational sciences. 
On a scale from 1 (no expertise) to 5 (high expertise) participants' average level of expertise was 3.24 in XR and 3.28 in 3D modeling.

\subsubsection{Procedure}
Upon following the link provided, participants were welcomed to a Qualtrics survey.
They provided consent and entered demographic data.
To lighten the time commitment, each participant was randomly assigned five locations (image pairs) out of the 10 total pairs and each image pair was evaluated on a Likert scale from 1 to 7 wrt.~four different categories: buildings, trees, terrain, and road network.
Each category was evaluated with two to four statement of the following nature: ``The architectural styles of buildings are accurately represented in  the virtual environment"; or ``The trees are realistic in terms of texture and appearance".
At the end of the survey, participants had to rate the suitability of the current realization wrt.~four different application domains: games, spatial planning, environmental data visualization, and stakeholder engagement.

\subsubsection{Methods}

To account for the repeated observations per participant, we employ both Frequentist and Bayesian mixed models~\cite{bates2015fitting,burkner2017brms}.
We approach model construction from two perspectives, see Eq.~\ref{eq:mixed:mod} and see Eq.~\ref{eq:mixed:mod:con}. 
Here $\beta$ are the coefficients for the fixed effects and $u$ are the coefficients for the random effects and $\epsilon$ is the random error.
We describe the model matrix in terms of column names rather than showing the full matrix in line with typical model forumla~\cite{wilkinson1973symbolic}.

First, we build mixed models that take location and kind of response into consideration as a vignette design, see Eq.~\ref{eq:mixed:mod}.
We assume that ratings across categories tap on the same visual fidelity construct. 
That is, each category is scored on the same overall visual fidelity spectrum, just focusing on separate aspects.
Therefore, we can model our survey experiment as a 10 locations times 4 types factorial design to obtain ratings from participants.
We let each participant observe 5 vignettes because it has been shown that repeated tests increase precision while yielding the same result as showing only one vignette per participant with more overall participants~\cite{CLIFFORD_SHEAGLEY_PISTON_2021}.
We obtain 1270 records of ratings for different experimental conditions and participants.
The interaction in Eq.~\ref{eq:model:fixed} is only used in model 1. 
The interaction in Eq.\ref{eq:model:random} is only used in model 2.

\begin{align}
    y &= X\beta + Zu + \epsilon \label{eq:mixed:mod}\\    
   \begin{split}
    X &= location \cdot type + gender + age + 3d\_experience + XR\_experience\\ \label{eq:model:fixed}
    &+ education + game\_developer + planner + visualisation\\ 
    &+ stakeholder\_engagement
    \end{split}\\
    Z &= participant + participant:type \label{eq:model:random}
\end{align}

Second, we build conjoint style models~\cite{hainmueller2015validating} where we observe objective differences between images (see Fig.~\ref{fig:examplesSurvey}).
We quantify how many trees (on a log scale) and objects are missing, whether buildings, roads or terrain looks noticeable and whether the perspective is aerial or from street-level (see Tab.~\ref{table:objective:measures}).
Instead of explicitly modeling locations, we assume the conjoint criteria allow us to qualify the differences, see Eq.~\ref{eq:mixed:mod:con}.
We use demographics as additional covariates.
In addition to a Frequentist model (model 3), we also build a Bayesian model (model 4).
We conduct a sensitivity analysis towards the location characteristics based on marginal effects~\cite{arel2024how}.
Marginal effects are useful because we can observe the impact a specific variable has while keeping all other variables at their mean or mode.

\begin{align}
    y &= X\beta + Zu + \epsilon \label{eq:mixed:mod:con}\\    
   \begin{split}
    X &= \log\#trees\_missing + \#objects\_missing + \#roads\_noticeable \\
    &+ \#buildings\_noticeable + \#terrain\_noticeable + perspective\\
    &+ gender + age + 3d\_experience + XR\_experience+ education\\ \label{eq:model:fixed:con}
    & + game\_developer + planner + visualisation\\ 
    &+ stakeholder\_engagement
    \end{split}\\
    Z &= participant\label{eq:model:random:con}
\end{align}

\subsubsection{Results} First, we present  the aggregated data and some of the main findings.
In the aggregated scores, the four main categories (buildings, trees, terrain, and road) received an average visual fidelity score ranging from 3.1 for trees to 3.8 for the terrain (out of 7).
Comparing the different locations, there are some visual differences.
The highest scores across all four categories were obtained by a market square seen from above (see top image pair in Figure \ref{fig:examplesSurvey}, the lowest scores from a very iconic location in Rotterdam, the Cube Houses, shown at street-level (see image pair second from top in Figure \ref{fig:examplesSurvey}; 4.2 versus 2.4).
The usefulness of the virtual environments created by AnywhereXR for different domains were ranked in the following order (with average scores out of 100): 67 for games, 51 for spatial planning, 51 for stakeholder engagement, and 48 for visualizing environmental data.

\begin{figure}[thb!]
\centering
\includegraphics[width=\columnwidth]{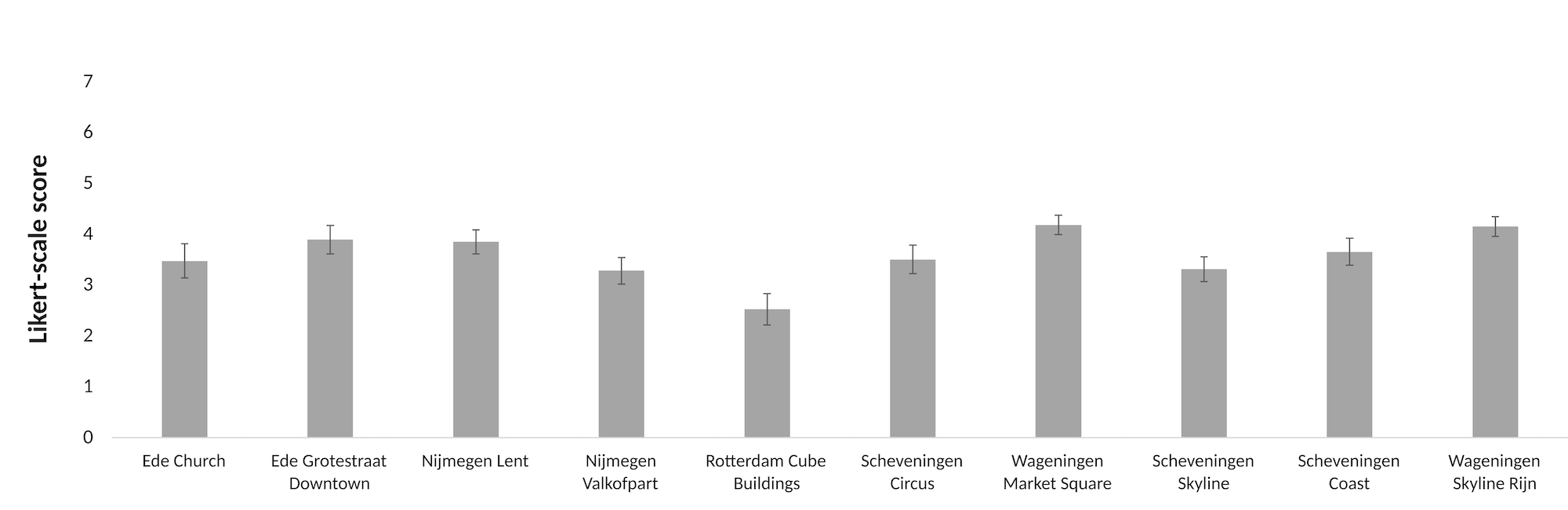}
\includegraphics[width=\columnwidth]{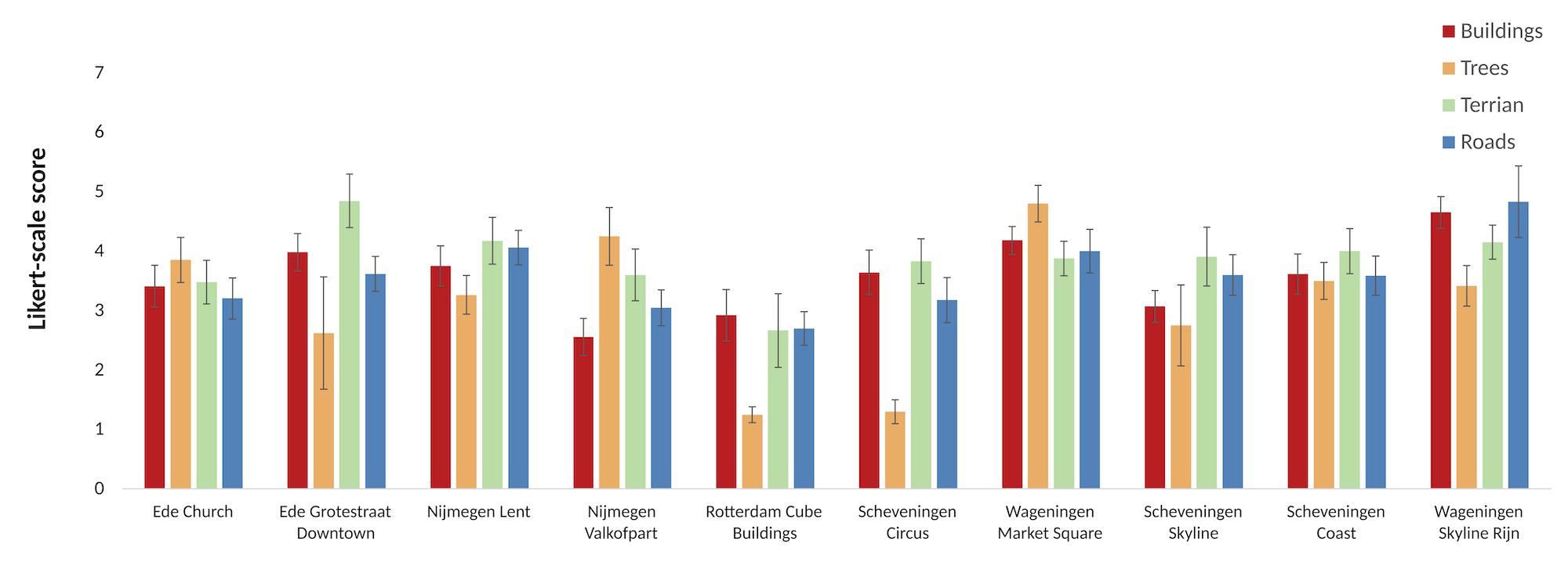}
\caption{Depicted are results from the survey. At the top, aggregated scores for all 10 image pairs; at the bottom, values for the four categories for each location.}\label{fig:surveyVis}
\end{figure}

\begin{figure}[thb!]
\centering
{\scriptsize
a)\includegraphics[width=\columnwidth]{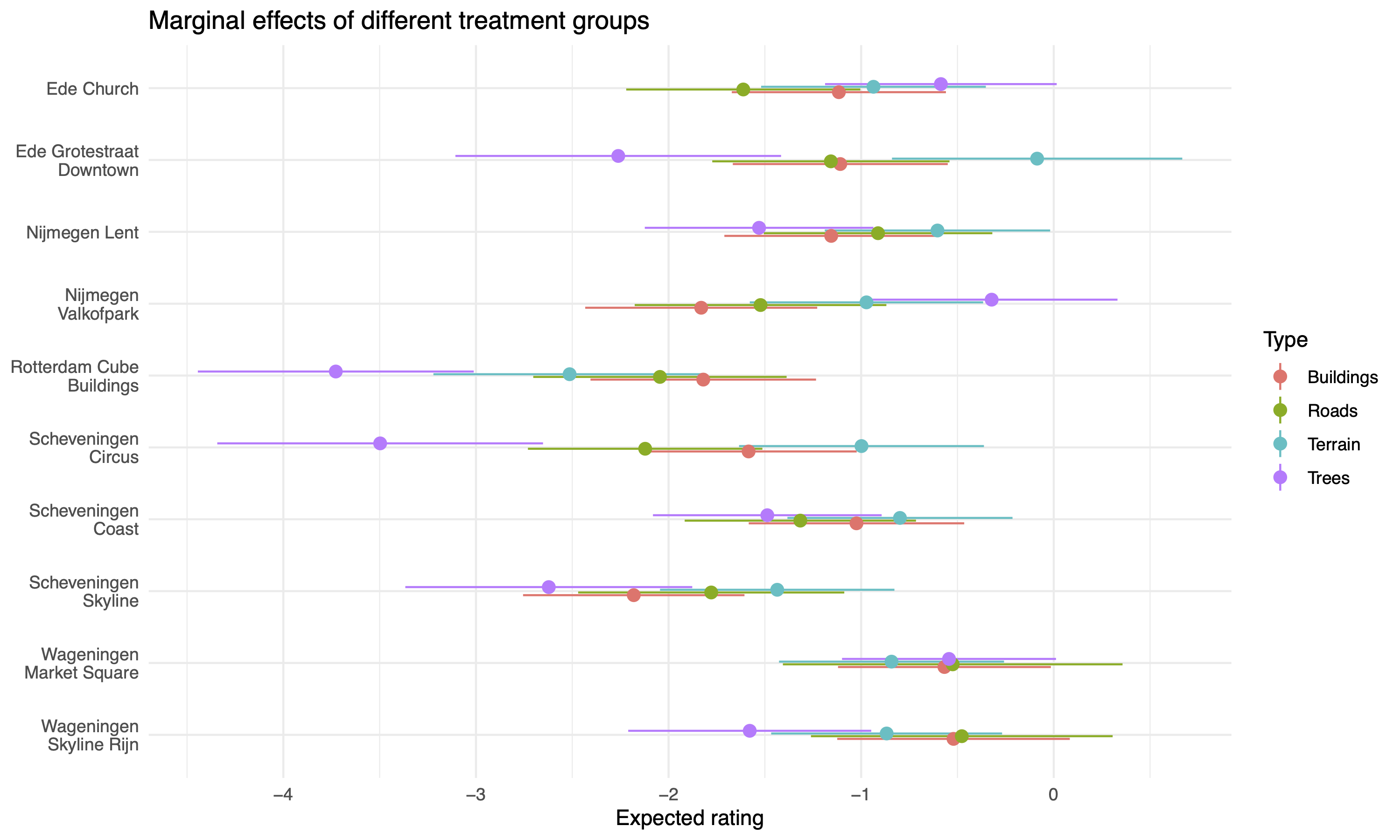}
b)\includegraphics[width=0.45\columnwidth]{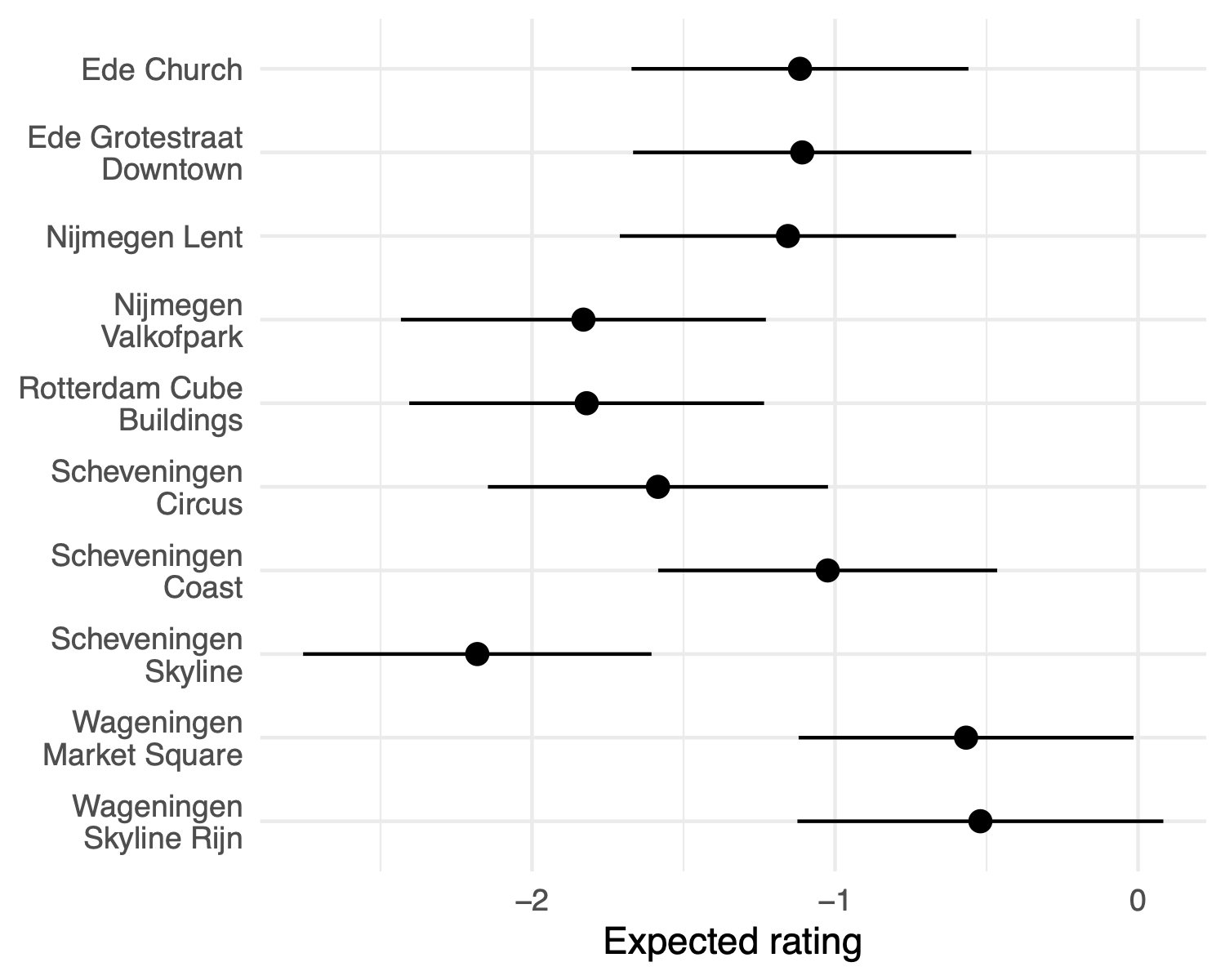}
c)\includegraphics[width=0.45\columnwidth]{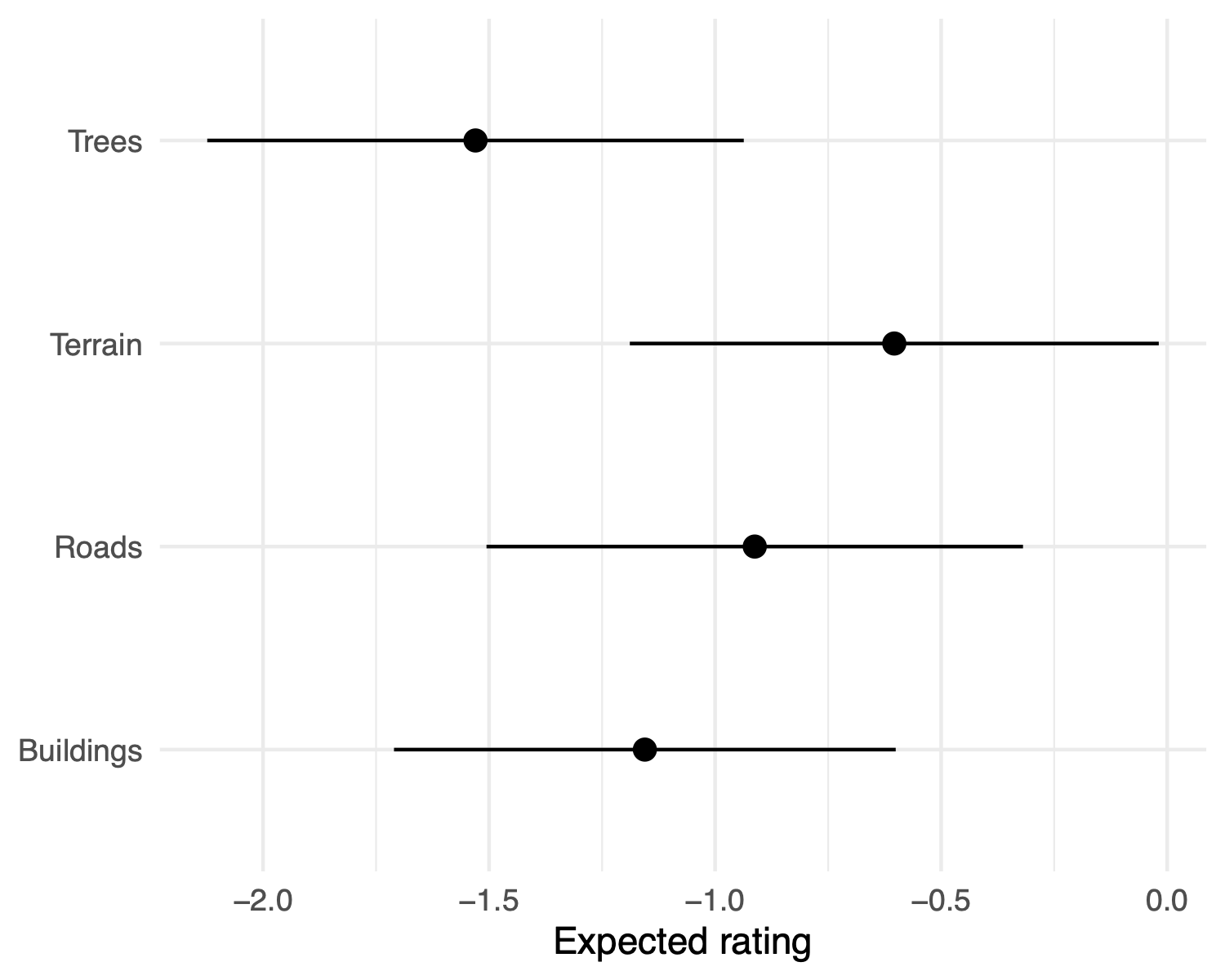}}
\caption{The marginal effect plots of the model per factorial (a). We also look at the marginal effect per location (b). We also look at the marginal effect of the question types (c). We find that trees play a particular role in deteriorating the perception.}\label{fig:survey:mod1}
\end{figure}

Our four models that explore our results exhibit consistency across all models (see Tab.~\ref{table:coefficients} in Appendix).
First, we evaluate the vignette design with marginal effects (see Fig.~\ref{fig:survey:mod1}a) on model 1.
With Ede Church and roads as the base-line scene and category, most scenes are rated significantly worse, except the Wageningen scenes which involve a lot of natural landscape and are rated significantly better, see Fig.~\ref{fig:survey:mod1}b.
Ratings are especially low at the Rotterdam Cube Building, the Nijmegen Valkofpark, and the Schevingen Skyline.
Schevingen Skyline performs worse than expected given the overall category scores. This is probably a reflection of the score being slightly lower across the board instead of being driven by tree ratings.
Our covariates do not have a significant effect.
From the interaction effects, we find that mostly the trees negatively impact ratings, see Fig.~\ref{fig:survey:mod1}c.
A simplified model without interaction and aggregated type results comes to mostly similar conclusions (model 2).

\begin{figure}[thb!]
\centering
{\scriptsize
a)\includegraphics[width=0.45\columnwidth]{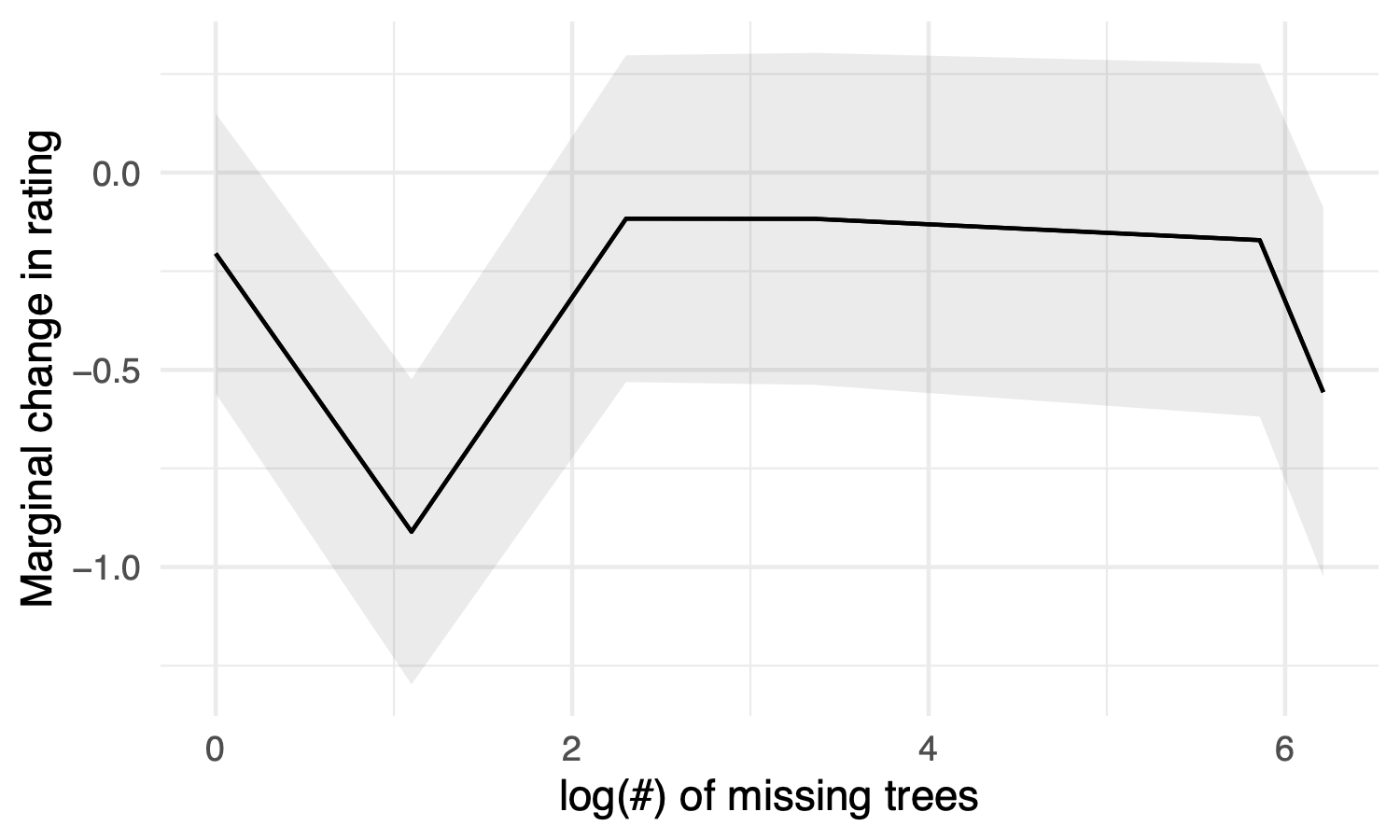}
b)\includegraphics[width=0.45\columnwidth]{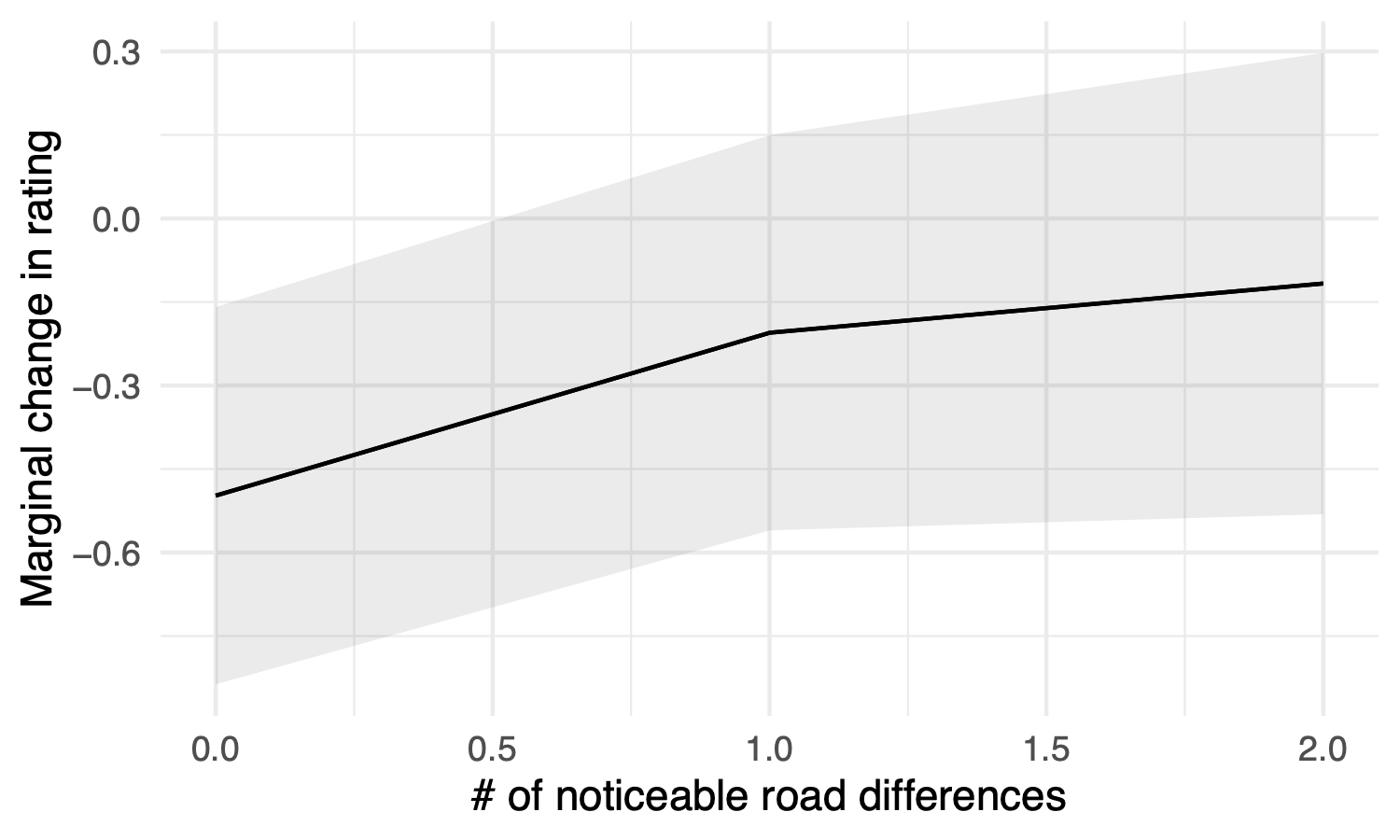}
c)\includegraphics[width=0.45\columnwidth]{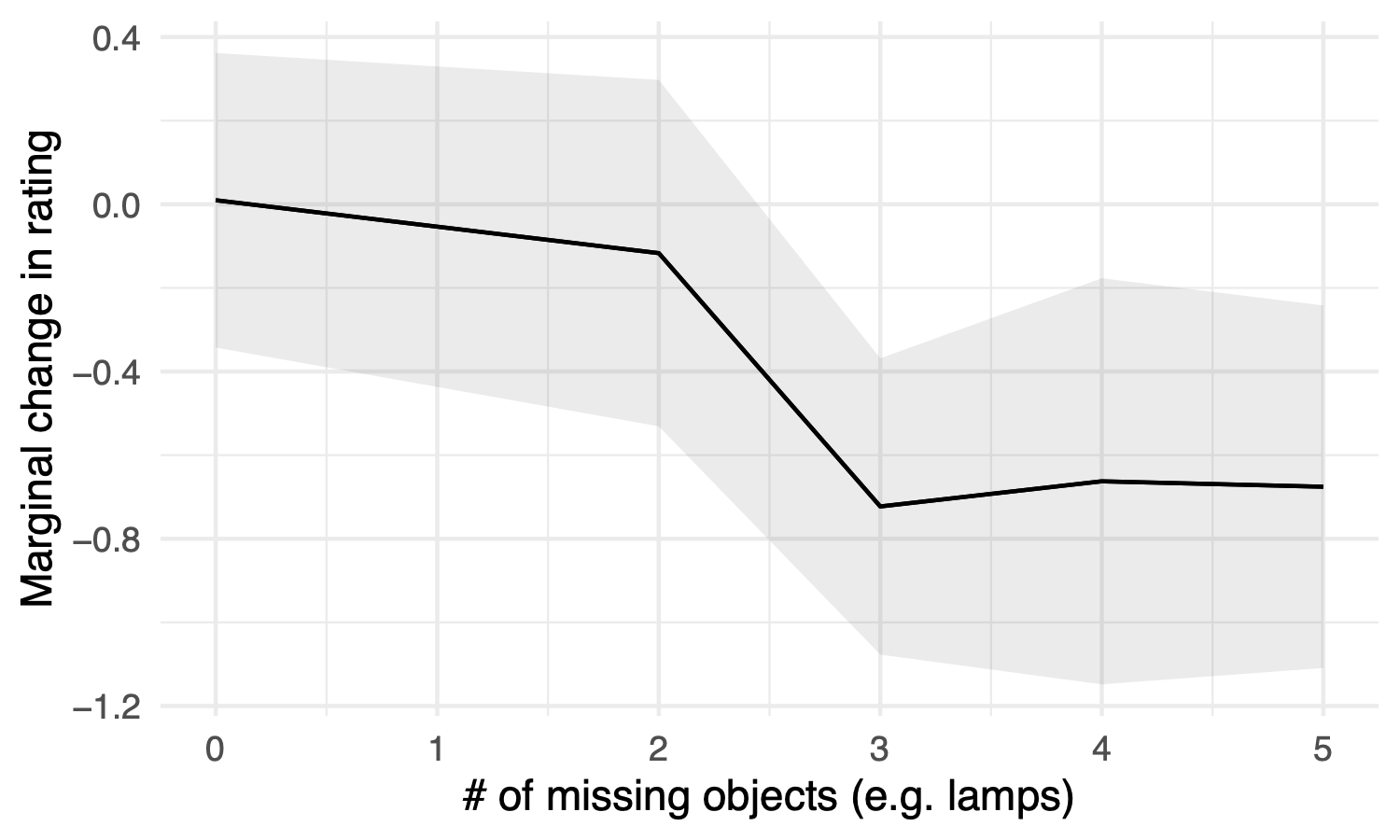}
d)\includegraphics[width=0.45\columnwidth]{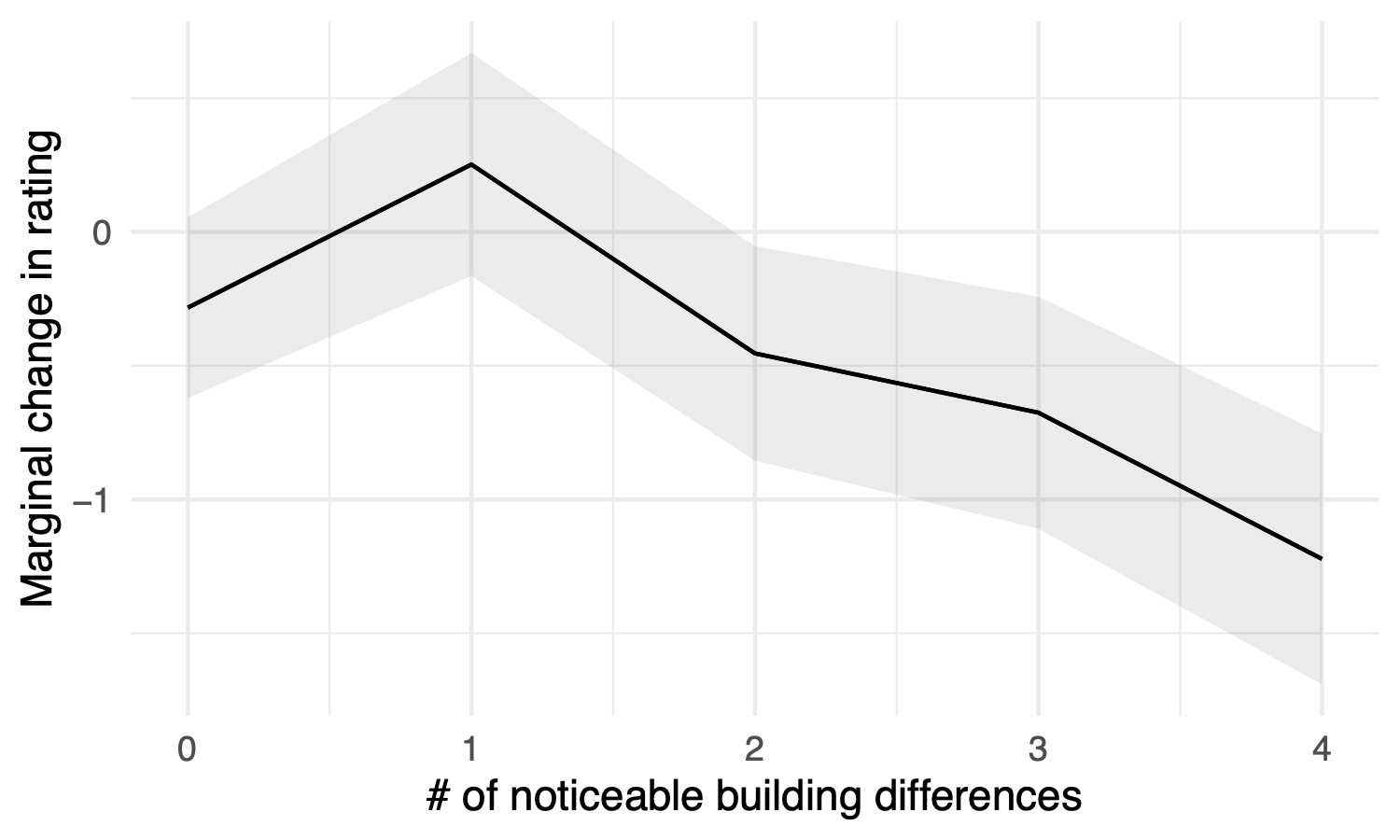}
e)\includegraphics[width=0.45\columnwidth]{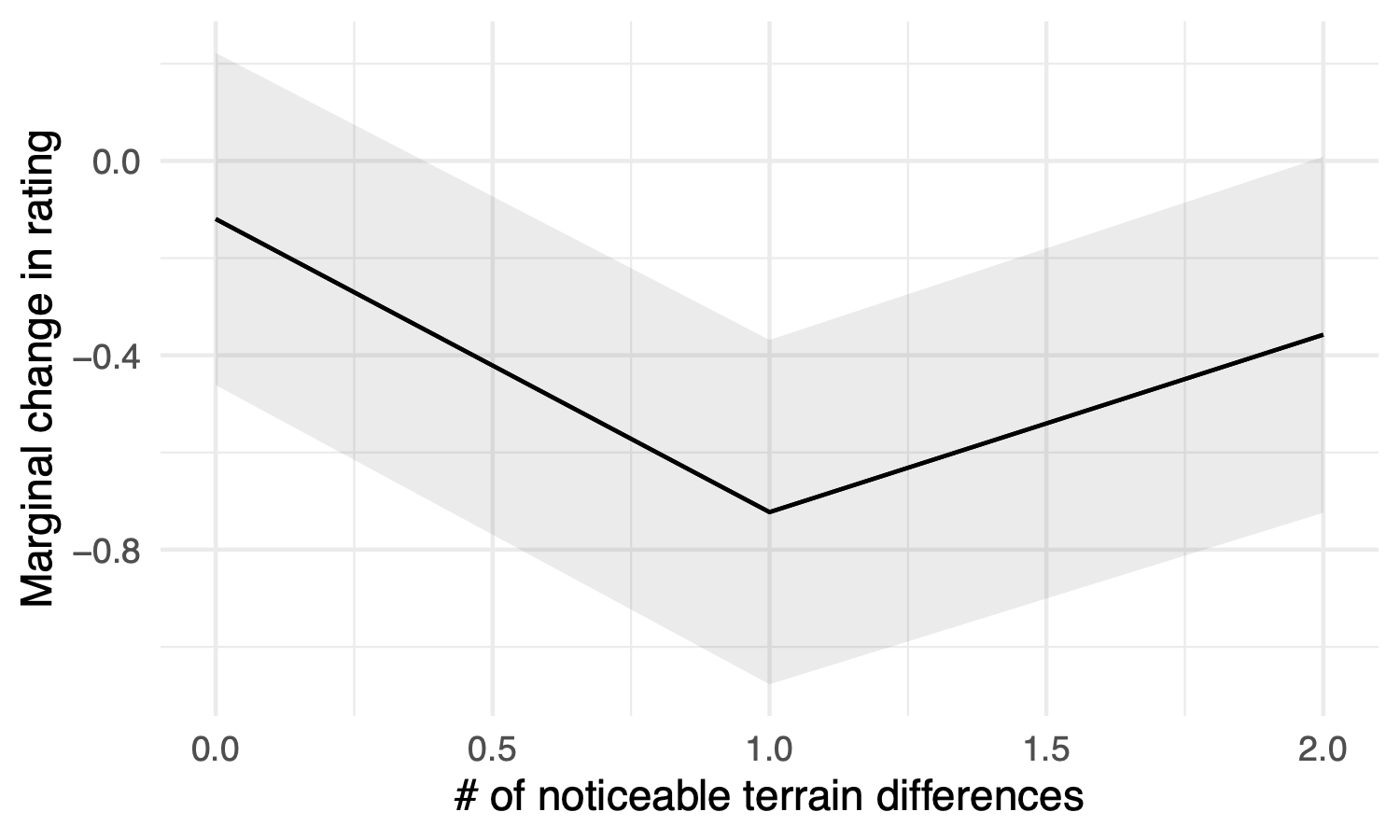}
f)\includegraphics[width=0.45\columnwidth]{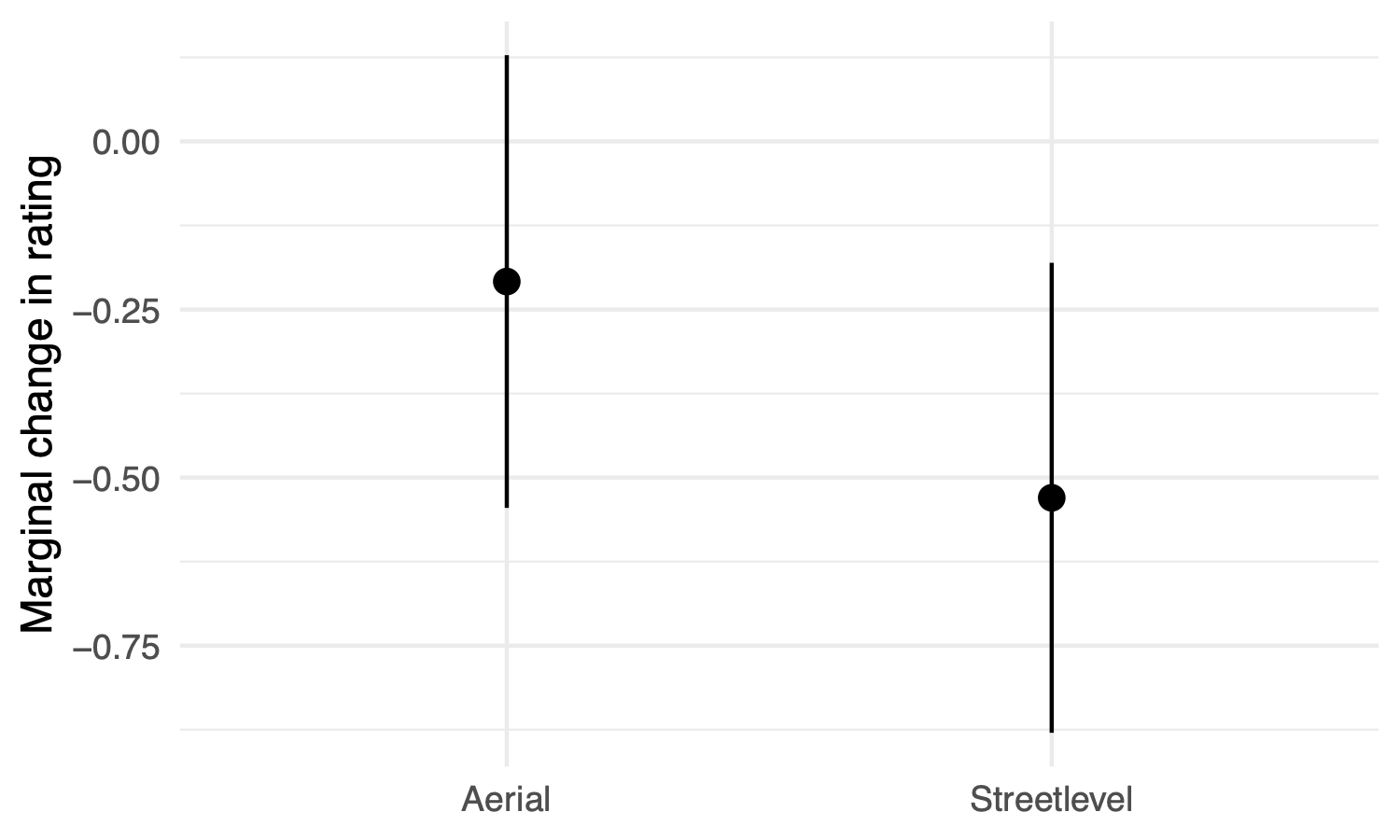}}
\caption{Marginal change plots of our conjoint model. We plot the sensitivity of our model variables as the expected rating across the range of the variable. We find that a single missing tree significantly deteriorates ratings (a). Multiple missing objects also deteriorate ratings (c). Weird buildings in larger numbers deteriorate ratings (d) but terrain and roads has little to no effect (e, b). Street-level views also reduce ratings (f).}\label{fig:survey:mod3}
\end{figure}

A sensitivity analysis based on marginal effects~\cite{arel2024how} was conducted and we find that people strongly react on a single tree missing, multiple objects missing, more than one strange-looking building, and closeness to surface of the observer (see Fig.~\ref{fig:survey:mod3} and model 3 in Tab.~\ref{table:coefficients}).

While several missing trees (on a log scale) do not strongly worsen the rating, it is a single missing tree that disturbs raters.
This was the case in the Rotterdam Cube Building picture, see Fig.~\ref{fig:survey:mod3}a.
It appears that characteristic trees in an image play a larger role than a whole forest.

We did not model smaller objects but some scenes contain them as crucial elements. When the number of objects is small enough, this is not considered negative but with more objects missing, it impacts the rating. 
This was the case in the Nijmegen Valkofpark, see Fig.~\ref{fig:survey:mod3}c.

A single building with an unusual noticeable look (texture) is not disturbing but multiple strange buildings seem to reduce the rating as for the Rotterdam Cube Building, see Fig.~\ref{fig:survey:mod3}d.
Lastly, the street-level perspective is rated lower than the aerial perspective, see Fig.~\ref{fig:survey:mod3}f.
Roads and terrain do not impact the ratings but the bright color of the beach in the Schevingen Skyline  drives the negative impact for a single noticeable terrain, see Fig.~\ref{fig:survey:mod3}e.
We validated our findings with a Bayesian model to confirm the robustness of our model and come to similar effects (model 4), see Tab.~\ref{table:coefficients}.

\subsubsection{Discussion} 
The survey, intended as a first evaluation of AnywhereXR, provided some interesting results and valuable insights.
It needs to be stressed that some of the results could have also been obtained from expert feedback (including the authors).
Challenges with ``Trees", for example, stem largely from locations with fewer trees in urban area.
There are some mismatches between physical reality (as per the image used) and the data derived from the database.
So a follow up assessment should use recent images taken by the authors and separate two aspects: the completeness of databases and the quality/fidelity of the 3D visualization of the database content.
Additionally, other elements of the green infrastructure such as hedges are not present in the database and probably added to the low scores for trees.
It needs to be stressed that AnywhereXR is a fully automated system with no post processing.
Comparing the results to actual images rather than evaluating the virtual environments in particular use cases provides additional challenges as expectations are higher (see discussion below).
Nonetheless, we see the overall results as encouraging to continue the AnywhereXR efforts.

One of the main reasons behind AnywhereXR is, however, to address challenges that other large-scale 3D visualization efforts have, that is, they look good from the distance but fail to deliver sufficient quality at street-level.
While we are in the position to use AnywhereXR as a starting point for creating street-level immersive experiences, the approach is not perfect yet and needs further improvement.

While the visual Turing test might be an interesting approach in and of itself, the use of virtual reality is often different.
Only few scenarios require the direct comparison of the real world with a digital representation of it, and in many cases, users of virtual worlds might look for a more context specific assessment of the usefulness beyond focusing on visual fidelity alone~\cite{jep2014403669}.
This can be shown in studies on fidelity that found that plausibility is more important than fidelity~\cite{Huang.2020,Huang.2021}.
Interactivity is also widely acknowledged as essential for immersive experiences~\cite{Park.2020,Yim.2017}.
In a recent study, Dane et al.~\cite{Dane.inpress} showed that a representation with objectively lower fidelity (e.g., fewer textures, lower resolution) than our examples received overwhelming praise from participants of a participatory co-design workshop.
The authors point out that several participants, especially from the municipalities, preferred a low fidelity environment to convey a certain level of uncertainty about the final designs.

Hence, while we obtained first valuable insights from our survey, we also see clear limitations of our approach.
We do believe though that the overall direction of AnywhereXR should be to create object-based representations as close to reality as allowed by the data and technology.
That said, we also would like to stress that for many applications passing the visual Turing test should not be a requirement for creating successful immersive experiences.

\section{Case Study: Live Public Transport Data}
\label{sec:casestudy}

In addition to the evaluation in the previous section, we want to demonstrate the modularity of AnywhereXR.
To that end, we provide a proof-of-concept that includes real-time data in our immersive digital twin approach. To this end, we integrated information from NDOV (national database in the Netherlands for public transport) into the IDT focusing on public transport, encompassing trams and buses. 
We selected this data due to our efforts to focus largely on publicly accessible data sets and transportation data’s accessibility through free querying.
This data is considered real-time as vehicles such as buses emit GPS locations at intervals of approximately one to two minutes.
However, this sampling frequency is insufficient for animating public transport within the IDT, as a simplistic interpolation between points would result in vehicles traversing through obstacles such as buildings and trees, rendering it impractical. 

\begin{figure}[thbp!]
\centering
\includegraphics[width=\columnwidth]{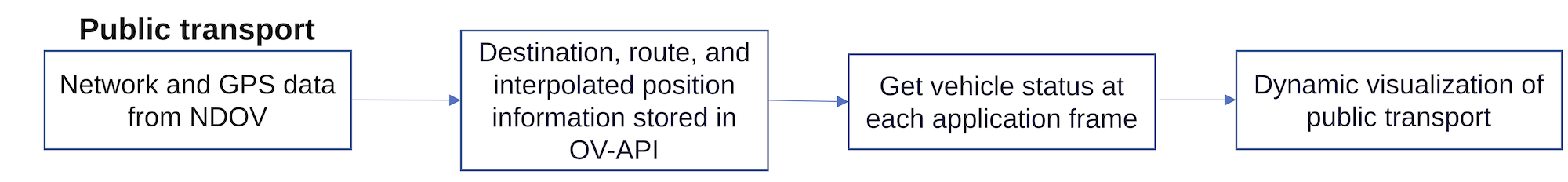}
\caption{Workflow for public transport in AnywhereXR.
}\label{fig:workflow_transportation}
\end{figure}

The integration of mobility simulations~\cite{strc2023chmove,lopez2018microscopic} would allow us, however, to place buses on roads in a future version and explore changes in public transport.
To effectively animate public transport vehicles within the environment, it becomes necessary to possess comprehensive knowledge of the entire transport network (Fig.~\ref{fig:workflow_transportation}).
This enables the extrapolation of vehicle positions along their designated routes into the future until updated GPS points become available.
However, the implementation of this feature presents significant challenges as it requires the aggregation of network data from all relevant companies and its synchronization with active public transport within a specific area.
Fortunately, in the Netherlands, this data is readily accessible and free for use, encoded within a specification known as NeTEx.\footnote{\url{https://netex-cen.eu/}}.

\subsection{NDOV Data}

NDOV stands for Nationale Data Openbaar Vervoer (EN: National Public Transport Data).
This database, maintained by the Dutch government, provides open access to a variety of public transportation data such as routes, schedules, and arrivals. 
Important for our purposes is that NDOV also provides pseudo-real-time data about the current location of buses, trams, trains, and ferries.

\subsection{OV-API}

The primary objective of our implementation is to track the position of vehicles, display their destinations, and illustrate their routes. 
To achieve this, a separate service, referred to as the OV-API, was developed independently from the landscape generator. 
The OV-API is responsible for reading the network data, incorporating live updates, interpolating vehicle positions along their routes, and transmitting this information via a socket connection (TCP/IP) to the landscape generator application upon request.
The landscape generator only receives a consolidated list of vehicles, each annotated with their respective position, destination, route, and additional relevant information. 
The OV-API serves as an intermediary, facilitating communication between the landscape generator and the cloud-based network data, thereby ensuring seamless integration of real-time public transport information into the digital twin environment and other potential applications. 
Any application can connect to the OV-API and request live updates.

\begin{figure}[t]
\tiny
\centering
(a) \includegraphics[height=3.52cm]{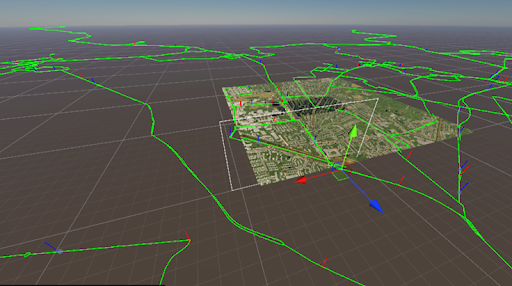}
(b) \includegraphics[height=3.52cm]{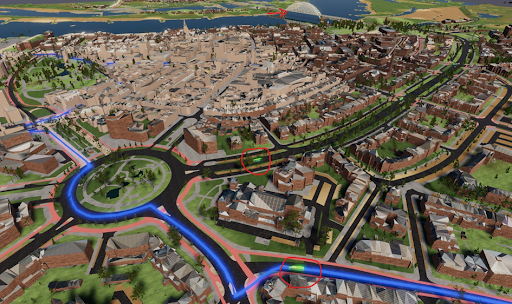}

(c) \includegraphics[height=3.52cm,trim={0 0 1.1cm 0},clip]{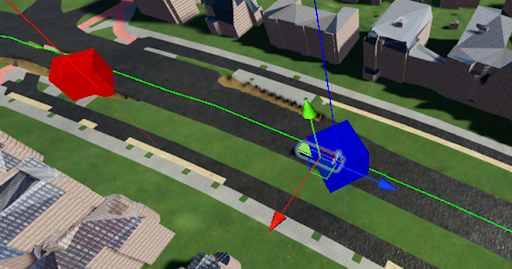}
\hspace{6.35cm}
\caption{(a) Visualization of the bus transportation network data provided by OV-API for the Nijmegen area inside the Unity editor.
(b) Visualization of vehicles and a selected bus route in the running IDT.
(c) Close-up view of a vehicle in the editor: The blue box shows the predicted position along its route, while the red one shows the position from the last known update.	
}\label{fig:transportation}
\end{figure}

Since the OV-API is only queried once per second, which is much lower than the frame rate of the application (e.g., 60 frames per second), the missing frames must be filled in. 
To do so, the data retrieved from the OV-API is stored in a buffer with a timestamp. 
At each frame, the application queries the buffer from the current application time with the traffic status estimated from the latest data of OV-API. 
Fig.~\ref{fig:transportation} shows different visualizations of the public transport information provided by OV-API.

\section{General Discussion}
\label{sec:gendisc}

In this paper, we combined three important developments: the growing availability of geospatial data, the maturation of XR technologies and supporting software environments, and opportunities related to digital twinning. 
There are a number of key characteristics of our approach that are worth noting. 
First, while not entirely achievable, our developments advance Open Research Data and Infrastructure by relying to a large degree on open, publicly accessible data and software environments. 
We firmly believe that in order to successfully address societal challenges at scale, the availability of high-quality, public data and Open Research Data practices are essential. 
We would like to praise the developments in the Netherlands that make data such as 3D BAG freely available. 
In some cases, such as the Bomenregister, the data is available in principle, but the full extent is hidden behind a paywall. 
Especially for efforts to monitor, communicate and make participatory decisions on biodiversity and climate-resilient planning, it will be essential to further increase the accessibility and public availability of such data sets~\cite{Coetzee.2020,Gomes.2020}. 
Another aspect is the integration of more diverse data sets.
Our efforts so far are almost exclusively based on authoritative data sets.
While these datasets offer exceptional quality and standardization, they are also slow to update and operate within quality restrictions. 
As seen by many citizen science projects 
(e.g., OSM\footnote{\url{https://www.openstreetmap.org/}}), there are many opportunities and advantages to also include non-authoritative data sets and engage the public in data collection and sharing~\cite{Vohland.2021,Fritz.2017,Sui.2013}.
Including multiple sources for the same information will also require resolution mechanisms for divergent data.
Future iterations of AnywhereXR will address both the visualisation of differences in sources and transparent resolution mechanisms.

Second, the developments rapidly change in all areas associated with this project, including the use of generative AI to control the looks of immersive environments~\cite{Ratican.2023}. 
It will be essential to look not only into the status quo and address its challenges but also look ahead.
In comparison to some previous approaches~\cite{Edler.2018}, the levels of realism and fidelity we reach based on open-source data are substantially advanced, yet not perfect.
The limitations of our approach are highlighted in the high quality data sources such as available for The Netherlands.
On the one hand, we are reliant on data quality to support high visual fidelity.
When data quality diminishes, our approach could result in a limited sense of presence. On the other hand, even the high data quality in The Netherlands is insufficient for the last mile of visual fidelity of adding small local objects (garbage bins, graffiti, hedges, parked vehicles, etc).
A promising AI approach based on generative simulation~\cite{xian2023towards} will be to enrich a scene in AnywhereXR with small local objects that similar locations in the real world would also exhibit. The information can be gathered from a wide range of geo-tagged video data sources and has already been gathered in repositories~\cite{mavros2023attenuating}. 
A LLM-based system can describe the contents of these videos~\cite{weng2024longvlm} to extract the types of small local objects in a scene.
A generative simulation system~\cite{xian2023towards} could create instructions for AnywhereXR to place small local objects found in the scene in credible locations.
Users of AnywhereXR would experience increased realism~\cite{rs14133095}.
The integration of AI for immersive experiences is carried out in various projects, and it is important to consider what the big picture should look like and how to, for example, create a metaverse / multiverse~\cite{Mystakidis.2022} infrastructure that complements industrial efforts~\cite{Rodriguez.2021} and European research networks such as the
European Metaverse Research Network\footnote{\url{https://metaversechair.ua.es/research-network/}}.

Third, one main advantage of our procedural landscape generation approach over 3D tile approaches (Tab.~\ref{tab:comparison}) used by Cesium or the 
Google Maps Tile API\footnote{\url{https://developers.google.com/maps/documentation/tile}} is that it creates a more detailed and realistic model at all scales, including the close-up ground-level perspective.
3D tile approaches tend to look blurry and sketchy at ground-level (Fig.~\ref{fig:3dtile}) because they are reconstructed from aerial data collections rather than high-quality open data.
This reconstruction can also impacts the soil representation and produces washed out images close up. 
In contrast, our  method incorporates soil information into a texture, which we use as an X by X grid to dynamically increase points and triangles based on the viewpoint.
For each grid point (pixel), the soil type is determined by referencing the texture.
The texture is employed as a constant source of information to dynamically generate terrain.
Our technique allows for the dynamic scaling of terrain complexity based on the viewpoint, significantly enhancing performance while preserving quality.
This is in contrast to earlier methods where triangular meshes derived from vector tiles are fixed in size.
These traditional methods often pre-generate terrain with multiple Levels of Detail (LoDs) to mitigate complexity issues and Cesium and Google map use LoD to reduce cost.
Our method allows for real-time adjustments in terrain complexity based on the viewpoint while benefiting from a constant texture size, offering greater flexibility during runtime adjustments.¨

\begin{table}[htb!]
    \centering
    \caption{Comparison of selected 3D mapping approaches}
    \begin{tabular}{lp{3cm}p{3cm}p{3cm}}
        \toprule
         & Aerial data capture & Vector tile & Dynamic vector tiles\\
         \midrule
         Examples& Vexcel, Google Maps & PDOK, Arcgis Online & AnywhereXR\\[3pt]
         Paradigm & Reconstructed & Object-oriented & Object-oriented\\[3pt]
         Scope & Continuous & Hierarchical & Local\\[2pt]
         Sourcing & 3D reconstruction & Rasterized vector tile & Open Research Data\\[3pt]
         Level of Detail & Pre-generated & Pre-generated & Dynamically sampled from texture\\[3pt]
         Fidelity & Medium & Low & High\\[2pt]
         \emph{Details}\\ 
         \phantom{a}street-level & Mushy shapes and textures &  Schematic & Authentic shapes, random textures\\[3pt]
         \phantom{a}Aerial view & Realistic & Schematic & Realistic\\[3pt]
         \emph{Objects} \\
         \phantom{a}Buildings & Irregular &  Untextured meshes & Authentic, random textures\\[3pt]
         \phantom{a}Trees & Blobs & - & Realistic\\[3pt]
         \phantom{a}Objects & Blobs & - & -\\
         Immersion potential & Low & None & High\\[2pt]
         \bottomrule
         \multicolumn{4}{r}{\tiny Note: "-" implies feature not implemented}
    \end{tabular}
    \label{tab:comparison}
\end{table}

Trees present another issue that is often insufficiently addressed.
Without precise tree maps, it becomes impossible to draw them well but aerial reconstructions get them for free.
In contrast, other smaller objects are often rendered as small blobs but may be hard to identify.
This is especially impactful for street-level visualizations. 
The potential for presence is driven by recognition and acceptance of the viewed scene. Whereas traditional vector tiles do not emphasize texture and are consequently impractical for immersive visualisation, the aerial data captures provide real-world information at the cost of blurred and jagged shapes.
Our approach provides both texturing and crisper shapes allowing users to more easily feel present in the environment.

\begin{figure}[thbp!]
\centering
\includegraphics[height=3.3cm]{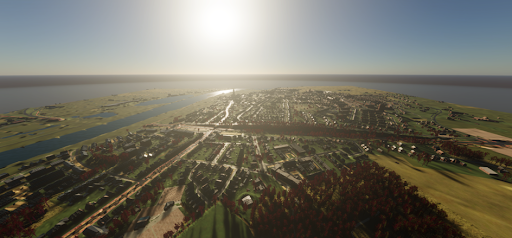}
\includegraphics[height=3.3cm]{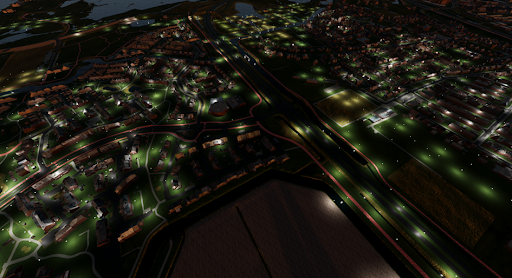}

\includegraphics[height=3.6cm]{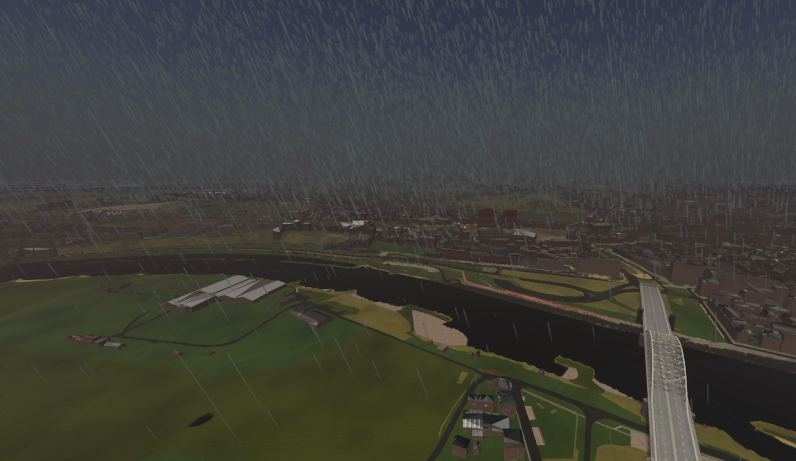}
\includegraphics[height=3.6cm]{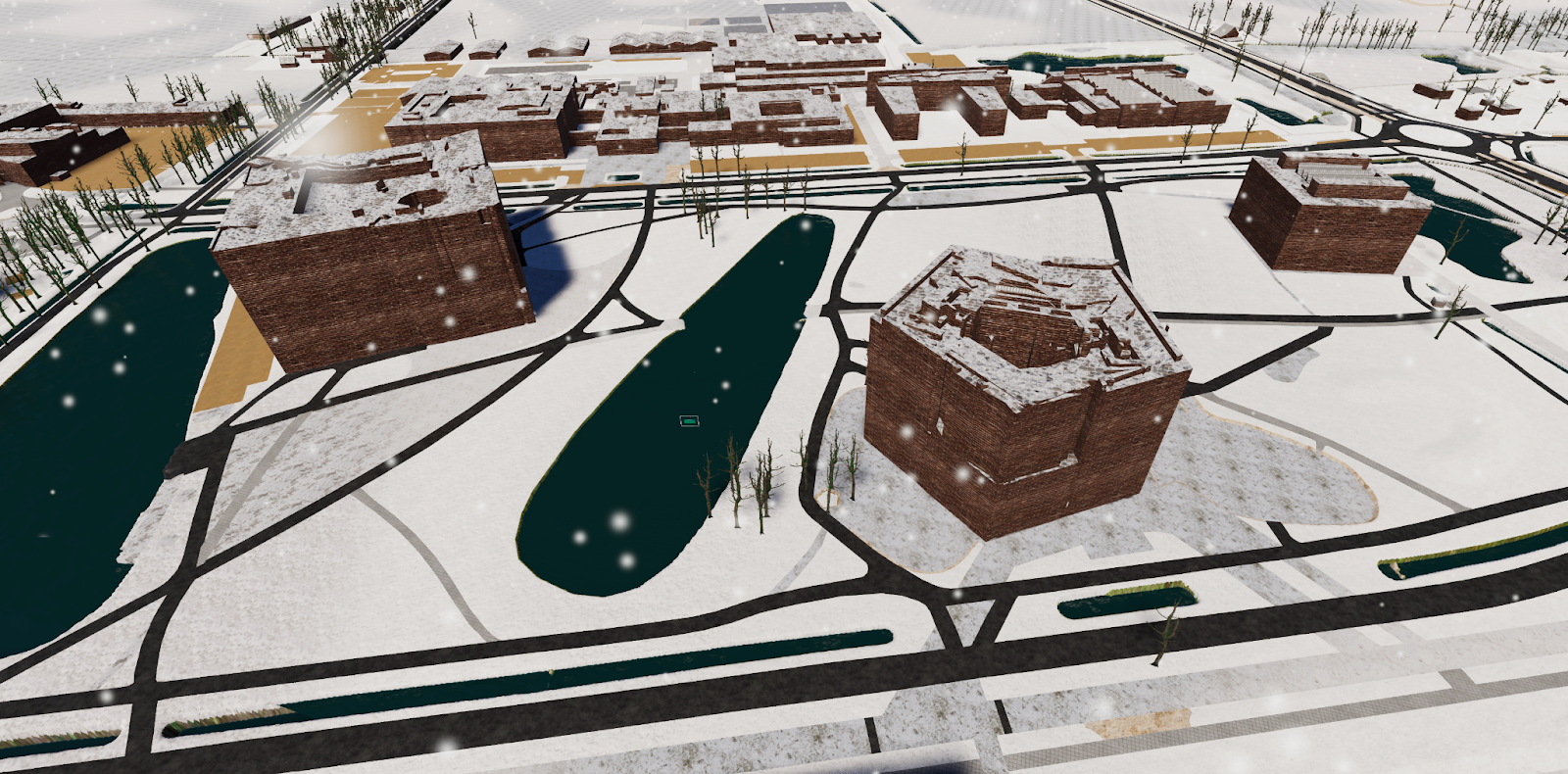}
\caption{Initial version of an IDT that incorporates day/night time changes (top) and weather effects like rain and snow (bottom).}\label{fig:timeandweather}
\end{figure}

Furthermore, our object-based representation is better suited to simulate the impacts of time of day/night time and weather as illustrated in Fig.~\ref{fig:timeandweather} or to adapt the environment to simulate alternative situations or scenarios.  
Currently, the weather simulations are not based on actual time and weather data, but live querying of weather stations will be one of the next steps. Regarding simulations of alternative scenarios, AnywhereXR, for instance, enables the creation and experience of potential future states of the modeled environment as a basis for informed decision-making. Intentional deviations from the actual state of the environment can be achieved by goal-directed modifications of the parameters of the model generation procedures at all stages, by integrating additional algorithms that modify the model after it has been created, or even by manual modifications of the generated model (e.g., in the game engine's scene editor). Examples of such applications could be the algorithmic simulation of the impacts of climate change on the future state of a vegetation, e.g., by varying the tree density and tree models based on input values stemming from an ecological forecast model~\cite{Huang.2020}, or the manual introduction of a change to the road network to evaluate different design options with different stakeholder groups. Exploring these options further will also be a next step on our agenda.

\section{Conclusions and Outlook}
\label{sec:conclusions}

The increasing availability of geo-spatial data, combined with advancements in digital twin technology, presents a significant opportunity to address societal and planetary challenges. 
However, to fully leverage this potential, it is essential to develop more intuitive, collaborative, and transparent methods of representing and interacting with data, information, and knowledge~\cite{Falco.2019,Thomas05:illuminating,Padilla.2018,Bauer.2024}. 
In this paper, we have introduced the concept of immersive digital twins (IDTs) within the framework of AnywhereXR. 
IDTs offer a human-centric approach to data representation, facilitating embodied, visceral data experiences for decision-making, including futuring, co-design, and co-creation.
In the following, we briefly discuss some key future developments.

\subsection{Collaborative IDT - From the Metaverse to the Multiverse}

A significant advancement in immersive technology is the realization of collaborative experiences~\cite{Pidel.2020}.
With the emergence of the Metaverse and related developments~\cite{Mystakidis.2022,Dwivedi.2023,Li.2024}, creating collaborative immersive experiences has become increasingly feasible. 
This shift has profound implications for addressing societal and planetary challenges, as collaborative approaches are crucial for processes such as participatory decision-making, co-design, and co-creation~\cite{SheppardStephenR.J..2015}. 
Integrating collaboration into IDTs can revolutionize stakeholder engagement, policy assessment, education, and research.

The fusion of digital twinning and immersive experiences opens up new possibilities for grounded communication and decision-making. 
Projects exploring immersive engagement in urban planning processes, for example, have demonstrated the potential benefits of this approach~\cite{Evers.2023,Ketzler.2020}.
Our work aims to democratize immersive methodologies on a larger scale, empowering also the public to participate in collaborative decision-making processes.

\subsection{Scalability}

We tackled scalability from two sides, computationally and data sourcing.
Previous approaches were often limited by high polygon counts to achieve high fidelity imagery.
In contrast, our approach uses dynamic polygon counts to maintain high fidelity close by while easing on computational costs for distant parts of the environment.
Although ambitious, our primary objective is to ensure that IDTs are available for locations worldwide.
Despite the challenges involved, the rapid advancements in spatial data infrastructure provide a clear direction for this endeavor and our prototype of the Netherlands provides an implementation that can be generalized to other data sources as well. 
At the European level, notable developments include initiatives like INSPIRE\footnote{[url{https://inspire.ec.europa.eu/}}, led by the European Commission. Although still in progress, participating countries have agreed upon a unified standard for geospatial data, aiming to establish a spatial data infrastructure that supports environmental policies and assessments. 
This is significant because spatial data has traditionally been managed independently by all countries. 
The existence of this open data infrastructure is already established, and the data we utilized for our project, such as PDOK, is nearly fully compatible with it.

An open aspect remains; how to handle scalability in the face of data scarcity? While it may be challenging to implement at scale, advancements of Gaussian splatting and improvement to recognize and identify object in created scenes might offer solutions at local scale~\cite{Wu.2024}. Other efforts that will fuel the opportunities for AnywhereXR approaches are developments in creating global scale data sets on soil and land cover~\cite{Safanelli.2025} or developments in global tree databases~\cite{Beech.2017}. 

\subsection{Further Integration of AI and DTs}

AnywhereXR serves as a foundational platform for creating immersive digital twins to enhance their contextual understanding through embodied experiences~\cite{rs14133095}.
As such, AnywhereXR is orthogonal to many Digital Twin initiatives and may be directly combined with them. 
For instance, integration with the Open Digital Twin Platform (ODTP) allows to split tasks between the operational aspects of digital twins (ODTP), the content management of digital twins (by users providing their analysis) and their communication through effective, immersive visualizations in AnywhereXR.
This will facilitate digital twin deployment on a global scale.

In the absence of data, proxies of the data may become useful. For instance, locations with high-quality data can be used to train AI algorithms~\cite{Wu.2024} that fill the gaps in data sparse but similar environments.
This idea can be expanded beyond data scarcity towards a level of detail that even the best data sources currently do not provide.
Here, advances in generative simulation are an outstanding opportunity to increase fidelity by filling in details that are not recorded but expected.
To leverage this potential, we propose integrating AI-based spatial interpolation techniques to infer missing elements (e.g., \cite{zhan2023ageneralized}), enabling the creation of realistic and coherent virtual environments even in data-sparse areas. 

Another area where AI applications will improve future immersive digital twins is interaction.
By leveraging AI advances in Large Language Models (LLM), we also aim to enable seamless interaction with digital twins, enhancing user engagement and accessibility. 
We are working towards language-driven immersive twins where you can instantiate a specific twin with a language command like ``Show Wageningen Campus in winter''.
Instead of configuring and picking parameters, the LLM can use knowledge-augmented generation~\cite{gan2024similarity,lewis2020retrieval} to operate AnywhereXR and setup an expected environment.

\subsection{Other Applications}

The concept of digital replicas holds promise across various domains, especially when focusing on digital humanism~\cite{werthner2020vienna} and addressing the replication crisis through increased reproducibility~\cite{aguilar2024experiments}. 
From addressing environmental crises to optimizing crowd management in dynamic settings, the applications are diverse and far-reaching and enable human-centric approaches where technology is not only applied for itself but towards a larger human goal~\cite{prem2024principles}. 
Immersive digital twins can also double experimentation frameworks~\cite{grubel2024experiments} and provide insights into basic research of how humans navigate, understand and manipulate the environment with experiments conducted in them~\cite{schinazi2023motivation}.
Leveraging landscape generation technology, diverse and realistic environments can be rapidly simulated, offering insights and solutions to multifaceted challenges with a focus on human perception, human engagement, and human values.

\section{Acknowledgements}
Part of this research is funded by NWO, the Dutch Research Council. Grant numbers 16869/OSF23.1.004, IVR-R, and, 184.036.008, Delta Enigma. The authors would like to thanks the WANDER XR Experience Lab at Wageningen University and Research for continuous support of this project.

%NWO (Dutch Research Council), and, Delta Enigmapen Science Fund 2023. NWO Delta Enigma. EWUU Take Part

\section{Ethics approval}
Ethics approval for the empirical study has been obtained from the authors' home institution.

\section{Data availability statement}
The newest source code for AnywhereXR is available on the gitlab of the university: \href{https://git.wur.nl/wander/anywherexr}{https://git.wur.nl/wander/anywherexr}. The persistent version of the source code can be found on zenodo: \href{https://doi.org/10.5281/zenodo.15019922}{https://doi.org/10.5281/zenodo.15019922}.

The evaluation data that support the findings of this study are also available on zenodo: \href{https://doi.org/10.5281/zenodo.15019922}{https://doi.org/10.5281/zenodo.15019922}.

\section{Disclosure of Interest}
There are no relevant financial or non-financial competing interests to report.

\bibliographystyle{elsarticle-num} 
\newpage
\bibliography{bibliography}

\appendix
\section{Regression models}

\begin{table}[h!]
\caption{Statistical models for analysis of survey responses}
\begin{center}
\resizebox{.74\textwidth}{!}{
\begin{tabular}{l c c c c}
\toprule
 & Model 1 & Model 2 & Model 3 & Model 4 \\
\midrule
Baseline (Ede Church, Building)                     & $-1.26 \; (1.20)$       & $-1.09 \; (1.18)$       & $-0.92 \; (1.22)$       & $0.70 \; [-0.18;  1.67]$    \\
Ede Grotestraat Downtown                      & $0.01 \; (0.18)$        & $-0.01 \; (0.12)$       &                         &                               \\
Neijmegen Lent                       & $-0.04 \; (0.18)$       & $0.02 \; (0.11)$        &                         &                               \\
Neijmegen Valkofpark                       & $-0.71 \; (0.22)^{**}$  & $-0.19 \; (0.13)$       &                         &                               \\
Rotterdam Cube Buildings                      & $-0.70 \; (0.20)^{***}$ & $-1.21 \; (0.13)^{***}$ &                         &                               \\
Schevingen Circus                  & $-0.47 \; (0.18)^{*}$   & $-0.74 \; (0.12)^{***}$ &                         &                               \\
Schevingen Coast                   & $0.09 \; (0.18)$        & $-0.11 \; (0.11)$       &                         &                               \\
Schevingen Skyline                       & $-1.06 \; (0.20)^{***}$ & $-0.98 \; (0.13)^{***}$ &                         &                               \\
Wageningen Market Square                      & $0.55 \; (0.18)^{**}$   & $0.44 \; (0.11)^{***}$  &                         &                               \\
Wageningen Skyline Rijn                      & $0.60 \; (0.22)^{**}$   & $0.10 \; (0.13)$        &                         &                               \\
Roads                        & $-0.50 \; (0.22)^{*}$   & $-0.08 \; (0.17)$       &                         &                               \\
Terrain                      & $0.18 \; (0.20)$        & $0.25 \; (0.17)$        &                         &                               \\
Trees                        & $0.53 \; (0.21)^{*}$    & $-0.24 \; (0.17)$       &                         &                               \\
Female                     & $0.73 \; (0.51)$        & $0.65 \; (0.50)$        & $0.76 \; (0.51)$        & $0.32 \; [-0.10;  0.70]$      \\
Age (in years)                              & $-0.01 \; (0.02)$       & $-0.02 \; (0.02)$       & $-0.01 \; (0.02)$       & $-0.00 \; [-0.02;  0.01]$     \\
3D modelling experience                         & $0.06 \; (0.20)$        & $0.06 \; (0.20)$        & $0.01 \; (0.20)$        & $0.02 \; [-0.13;  0.18]$      \\
XR experience                         & $0.03 \; (0.21)$        & $0.01 \; (0.21)$        & $0.01 \; (0.22)$        & $0.00 \; [-0.16;  0.17]$      \\
Doctorate Degree        & $0.91 \; (0.85)$        & $0.90 \; (0.84)$        & $0.94 \; (0.86)$        &  $0.33 \; [-0.34;  0.99]$   \\
Master's Degree         & $0.56 \; (0.80)$        & $0.61 \; (0.79)$        & $0.68 \; (0.81)$        &   $0.21 \; [-0.41;  0.84]$ \\
Proficient in Game Development                        & $0.00 \; (0.75)$        & $0.04 \; (0.74)$        & $-0.03 \; (0.76)$       & $0.05 \; [-0.52;  0.62]$      \\
Proficient in Spatial Planning                         & $0.04 \; (1.11)$        & $0.17 \; (1.09)$        & $0.47 \; (1.12)$        & $0.10 \; [-0.76;  0.94]$      \\
Proficient in Visualisation                       & $0.15 \; (1.61)$        & $-0.23 \; (1.59)$       & $-0.22 \; (1.62)$       & $0.03 \; [-1.22;  1.25]$      \\
Proficient in Stakeholder Engagement                      & $0.81 \; (1.05)$        & $1.01 \; (1.03)$        & $0.78 \; (1.06)$        & $0.24 \; [-0.54;  1.00]$      \\
Roads x Ede Grotestraat Downtown            & $0.45 \; (0.31)$        &                         &                         &                               \\
Roads x Neijmegen Lent          & $0.74 \; (0.30)^{*}$    &                         &                         &                               \\
Roads x Neijmegen Valkofpark             & $0.80 \; (0.35)^{*}$    &                         &                         &                               \\
Roads x Rotterdam Cube Buildings           & $0.27 \; (0.34)$        &                         &                         &                               \\
Roads x Schevingen Circus      & $-0.04 \; (0.31)$       &                         &                         &                               \\
Roads x Schevingen Coast        & $0.20 \; (0.31)$        &                         &                         &                               \\
Roads x Schevingen Skyline             & $0.90 \; (0.36)^{*}$    &                         &                         &                               \\
Roads x Wageningen Skyline Rijn           & $0.54 \; (0.41)$        &                         &                         &                               \\
Terrain x Ede Grotestraat Downtown          & $0.84 \; (0.37)^{*}$    &                         &                         &                               \\
Terrain x Neijmegen Lent                    & $0.37 \; (0.28)$        &                         &                         &                               \\
Terrain x Neijmegen Valkofpark              & $0.68 \; (0.32)^{*}$    &                         &                         &                               \\
Terrain x Rotterdam Cube Buildings          & $-0.87 \; (0.36)^{*}$   &                         &                         &                               \\
Terrain x Schevingen Circus            & $0.41 \; (0.31)$        &                         &                         &                               \\
Terrain x Schevingen Coast               & $0.05 \; (0.28)$        &                         &                         &                               \\
Terrain x Schevingen Skyline                & $0.57 \; (0.30)$        &                         &                         &                               \\
Terrain x Wageningen Market Square           & $-0.45 \; (0.28)$       &                         &                         &                               \\
Terrain x Wageningen Skyline Rijn           & $-0.53 \; (0.31)$       &                         &                         &                               \\
Terrain x Ede Grotestraat Downtown         & $-1.68 \; (0.43)^{***}$ &                         &                         &                               \\
Terrain x Neijmegen Lent                   & $-0.90 \; (0.30)^{**}$  &                         &                         &                               \\
Terrain x Neijmegen Valkofpark             & $0.98 \; (0.35)^{**}$   &                         &                         &                               \\
Terrain x Rotterdam Cube Buildings         & $-2.44 \; (0.37)^{***}$ &                         &                         &                               \\
Terrain x Schevingen Circus              & $-2.44 \; (0.43)^{***}$ &                         &                         &                               \\
Terrain x Schevingen Coast                 & $-0.99 \; (0.30)^{***}$ &                         &                         &                               \\
Terrain x Schevingen Skyline               & $-0.97 \; (0.38)^{*}$   &                         &                         &                               \\
Terrain x Wageningen Market Square         & $-0.51 \; (0.27)$       &                         &                         &                               \\
Terrain x Wageningen Skyline Rijn          & $-1.59 \; (0.34)^{***}$ &                         &                         &                               \\
Log Number of missing trees                   &                         &                         & $-0.11 \; (0.03)^{**}$  & $-0.04 \; [-0.07; -0.02]^{*}$ \\
Number of missing objects                    &                         &                         & $-0.12 \; (0.04)^{***}$ & $-0.04 \; [-0.06; -0.02]^{*}$ \\
Number of noticeable buildings                  &                         &                         & $-0.14 \; (0.06)^{*}$   & $-0.07 \; [-0.11; -0.03]^{*}$ \\
Number of noticeable roads                    &                         &                         & $0.19 \; (0.14)$        & $0.07 \; [-0.02;  0.16]$      \\
Number of noticeable terrain                  &                         &                         & $-0.07 \; (0.08)$       & $-0.02 \; [-0.08;  0.03]$     \\
street-level perspective          &                         &                         & $-0.31 \; (0.16)^{*}$   & $-0.13 \; [-0.24; -0.03]^{*}$ \\
\midrule
AIC                              & $3730.71$               & $3693.88$               & $4551.53$               &                            \\
BIC                              & $3993.20$               & $3827.70$               & $4647.98$               &                             \\
Log Likelihood                   & $-1814.36$              & $-1820.94$              & $-2256.76$              &                             \\
Num. obs.                        & $1270$                  & $1270$                  & $1184$                  & $1184$                        \\
Num. groups: ResponseId          & $28$                    & $28$                    & $28$                    &                             \\
Var: ResponseId (Intercept)      & $0.71$                  & $0.61$                  & $0.68$                  &                             \\
Var: Residual                    & $0.93$                  & $0.85$                  & $2.47$                  &                             \\
Num. groups: ResponseId:Type     &                       & $111$                   &                       &                             \\
Var: ResponseId:Type (Intercept) &                       & $0.32$                  &                       &                             \\
SD: ResponseId                   &                       &                      &                       & $0.39$                        \\
R$^2$                            &                       &                       &                       & $0.23$                        \\
loo IC                           &                       &                       &                       & $4394.26$                     \\
WAIC                             &                       &                       &                      & $4394.13$                     \\
\bottomrule
\multicolumn{5}{l}{\scriptsize{$^{***}p<0.001$; $^{**}p<0.01$; $^{*}p<0.05$ (or Null hypothesis value outside the confidence interval).}}
\end{tabular}
}
\label{table:coefficients}
\end{center}
\end{table}

\begin{table}[h!]
\caption{Objective measures on image differences between AnywhereXR and comparison images}
\centering
\resizebox{.97\textwidth}{!}{
\begin{tabular}{rlrrrrrl}
\toprule
 & Location & \makecell[c]{\#missing\\trees} & \makecell[c]{\#missing\\objects} & \makecell[c]{\#noticeable\\building} & \makecell[c]{\#noticeable\\roads} & \makecell[c]{\#noticeable\\terrains} & perspective \\ 
\midrule
1 & edc & 3.00 & 5.00 & 3.00 & 1.00 & 2.00 & street-level \\ 
  2 & egd & 0.00 & 3.00 & 2.00 & 1.00 & 1.00 & street-level \\ 
  3 & rcb & 1.10 & 3.00 & 4.00 & 0.00 & 1.00 & street-level \\ 
  4 & scircus & 1.10 & 5.00 & 3.00 & 0.00 & 2.00 & aerial \\ 
  5 & scoast & 3.37 & 0.00 & 0.00 & 0.00 & 2.00 & aerial \\ 
  6 & ss & 6.21 & 3.00 & 0.00 & 0.00 & 1.00 & aerial \\ 
  7 & wsr & 5.86 & 0.00 & 0.00 & 0.00 & 0.00 & aerial \\ 
  8 & nv & 0.00 & 4.00 & 0.00 & 1.00 & 0.00 & street-level \\ 
  9 & nl & 2.30 & 2.00 & 0.00 & 2.00 & 0.00 & street-level \\ 
  10 & wms & 0.00 & 0.00 & 1.00 & 1.00 & 0.00 & aerial \\
  \bottomrule
\end{tabular}
}
\label{table:objective:measures}
\end{table}



%\begin{thebibliography}{00}

%% For numbered reference style

%\end{thebibliography}
\end{document}